\documentstyle[aps,preprint,pra,epsf]{revtex}

\title{Entanglement Measures and Purification Procedures}
\author{V. Vedral and M.B. Plenio}
\address{Optics Section, Blackett Laboratory, Imperial College London, 
London SW7 2BZ, England}
\date{\today}

\begin{document}

\maketitle

\begin{abstract}
We improve previously proposed conditions each measure of 
entanglement has to satisfy. We present a class of entanglement 
measures that satisfy these conditions and show that the Quantum 
Relative Entropy and Bures Metric generate two measures of this 
class. We calculate the measures of entanglement for a number of
mixed two spin $1/2$ systems using the Quantum Relative Entropy, 
and provide an efficient numerical method to obtain the measures 
of entanglement in this case. In addition, we prove a number
of properties of our entanglement measure which have important
physical implications.
We briefly explain the statistical basis of our measure 
of entanglement in the case of the Quantum Relative Entropy. 
We then argue that our entanglement measure determines an upper bound  
to the number of singlets that can be obtained by any purification 
procedure.  
\end{abstract}
\noindent
PACS: 03.65.Bz

\section{Introduction}

It was thought until recently that
Bell's inequalities provided a good criterion for separating quantum
correlations (entanglement) from classical ones in a given 
quantum state. While it is true that a violation of Bell's
inequalities is a signature of quantum correlations (non--locality),
not all entangled states violate Bell's inequalities \cite{Gisin}.
So, in order to completely separate quantum from classical
correlations a new criterion was needed. This also initiated the
search into the  related question of the amount of entanglement
contained in a given quantum state.
There are a number of `good' measures of the amount of 
entanglement for two quantum systems in a pure state 
(see \cite{Ekert1} for an extensive presentation).  
A `good' measure of entanglement for mixed states is, 
however, very hard to find. 
In an important work Bennett et al \cite{Bennett1}
have recently proposed three measures of entanglement (we will discuss the
entanglement of formation and distillation in more detail later in this paper).
Their measures are based on concrete physical ideas and are
intuitively easy to understand. They investigated many properties of these measures and calculated the entanglement
of formation for a number of states.  More recently, Hill and Wootters
have proposed a closed form for the entanglement of formation for two spin $1/2$ particles \cite{Wootters}. Uhlmann's recent work 
implies that the entanglement of formation can also be calculated numerically
in an efficient way for those cases that are not analytically known \cite{Uhlmann}.

We have recently shown how to 
construct a whole class of measures of 
entanglement \cite{Vedral,Vedral2}, and also imposed 
conditions that any candidate for such a 
measure has to satisfy\cite{Vedral}.
In short, we
consider the disentangled states which form a  convex subset 
of the set of all quantum states. Entanglement is then 
defined as a distance (not necessarily in the mathematical sense)
from a given state to this subset of disentangled states (see Fig. 1).  
An attractive feature
of our measure is that it is independent of the number of
systems and their dimensionality, and is therefore completely
general, \cite{Vedral,Vedral2}.
We present here two 
candidates for measuring distances on our set of states
and prove that they satisfy improved 
conditions for a measure of entanglement 
(the third condition presented here is an improvement 
over the one given in \cite{Vedral}). 

It should be noted that in much
the same way we can calculate the amount of classical correlations in a 
state. One would then define another subset, namely that of all product
states which do not contain any classical correlations. Given a disentangled
state one would then look for the closest uncorrelated state. The distance
could be interpreted as a measure of classical correlations.
In addition to many analytical results we also explain how to 
calculate efficiently using numerical methods our measure of entanglement of 
two spin $1/2$ particles. We present a number of examples and prove
several properties of our measure which have important
physical consequences. To illuminate the physical meaning behind
the above ideas we present a statistical 
view of our entanglement measure in the case of the Quantum 
Relative Entropy \cite{Vedral2}. We then relate our measure to a 
purification procedure and use it to define a reversible 
purification. This reversible purification is then linked 
to the notion of entanglement through the idea of 
distinguishing two classes of quantum states. We also argue that 
the measure of entanglement generated by the Quantum Relative Entropy that we propose gives an upper bound for 
the number of singlet states that can be distilled from a given
state. We find that in general the distillable entanglement is smaller than
the entanglement of creation. This result was independently proven by Rains for 
Bell diagonal states using completely different methods \cite{Rains}.

The rest of the paper is organized as follows. Section II introduces
the basis of purification procedures, conditions for a measure of
entanglement and our suggestion for a measure of entanglement. We also
prove that the Quantum Relative Entropy and the Bures Metric satisfy the
imposed conditions and can therefore be used as generators 
of measures of entanglement. We compute our measure explicitly for 
some examples.
In Section III we introduce a simple numerical
method to compute our measure of entanglement numerically and we apply it
to the case of two spin $1/2$
systems. We present a number of examples of entanglement computations
using the Quantum Relative Entropy. In Section IV
we present a statistical basis for the Quantum Relative Entropy as
a measure of distinguishability between quantum states and hence 
of amount of entanglement. Based on this, in Section V
we derive an upper bound to the efficiency (number of maximally 
entangled pairs distilled) of any purification procedure. We also
show how to extend our measure to more than two subsystems.
 
\section{Theoretical Background}

\subsection{Purification Procedures}

There are three different ingredients involved in 
procedures aiming at distilling locally a subensemble 
of highly entangled states from an original ensemble 
of less entangled states.

\begin{enumerate}

\item {\em Local general measurements} (LGM): these are 
performed by the two parties $A$ and $B$ separately and 
are described by two sets of operators satisfying the 
completeness relations $\sum_i {A}^{\dagger}_i {A}_i = 
{{1 \hspace{-0.3ex} \rule{0.1ex}{1.52ex}\rule[-.01ex]{0.3ex}{0.1ex}}}$
and $\sum_j {B}^{\dagger}_j {B}_j = {{1 \hspace{-0.3ex} \rule{0.1ex}{1.52ex}\rule[-.01ex]{0.3ex}{0.1ex}}}$. 
The joint action of the two is
described by $\sum_{ij} A_i\otimes B_j = \sum_i A_i 
\otimes \sum_j B_j$, which is again a complete general 
measurement, and obviously local. 

\item {\em Classical communication} (CC): this means that 
the actions of $A$ and $B$ can be correlated. This can be 
described by a {\em complete measurement} on the whole space 
$A+B$ and is not necessarily decomposable into a sum of 
direct products of individual operators (as in LGM). 
If ${\rho}_{AB}$ describes the initial state shared between 
$A$ and $B$ then the transformation involving `LGM+CC' 
would look like

\begin{equation}
\Phi(\rho_{AB}) = \sum_i  A_i \otimes  B_i 
\; {\rho}_{AB} \; A^{\dagger}_i \otimes  B^{\dagger}_i \;\; ,
\label{1i}
\end{equation}
where $\sum_i A^{\dagger}_iA_i B^{\dagger}_i B_i= {{1 \hspace{-0.3ex} \rule{0.1ex}{1.52ex}\rule[-.01ex]{0.3ex}{0.1ex}}}$, 
i.e. the actions of $A$ and $B$ are `correlated'.

\item {\em Post-selection} (PS) is performed on the 
{\em final ensemble} according to
the above two procedures. Mathematically this amounts 
to the general
measurement not being complete, i.e. we leave out some 
operations. The
density matrix describing the newly obtained ensemble 
(the subensemble of the original one) has to be 
renormalized accordingly. Suppose that we kept only
the pairs where we had an outcome corresponding to 
the operators $A_i$ and $B_j$, then the state 
of the chosen subensemble would be 

\begin{equation}
\rho_{AB} \longrightarrow   \frac{ A_i \otimes  B_i
\; {\rho}_{AB} \; A^{\dagger}_i \otimes  B^{\dagger}_i}
{\mbox{Tr} ( A_i \otimes  B_i \; {\rho}_{AB} \;  
A^{\dagger}_i \otimes  B^{\dagger}_i)}
\label{2i}
\end{equation}
where the denominator provides the necessary normalization.  

\end{enumerate}

\noindent
A manipulation involving any of the above three 
elements or their combination we shall henceforth call 
a {\em purification procedure}. It should be noted 
that the three operations described above are local. 
This implies that the entanglement of the total ensemble 
cannot increase under these operations. However, classical 
correlations between the two subsystems {\em can} be increased, 
even for the whole ensemble, if we allow classical communication. 
A simple example confirms this. Suppose that the initial ensemble 
contains states 
$|0_A\rangle \otimes (|0_B\rangle + |1_B\rangle)/\sqrt{2}$. 
The correlations (measured by e.g. von Neumann's mutual 
information \cite{Ekert1,Vedral}) between A and B are zero. 
Suppose that B performs 
measurement of his particles in the standard $0$, $1$
basis. If $1$ is obtained, B communicates this to A who then 
``rotates" his qubit to the state $|1_A\rangle$. 
Otherwise they do nothing. The final state
will therefore be
\begin{equation}
\rho = \frac{1}{2}(|0_A\rangle \langle 0_A | \otimes 
|0_B\rangle \langle 0_B |+ |1_A\rangle \langle 1_A | 
\otimes |1_B \rangle \langle 1_B |) \;\; ,
\end{equation}
where the correlations are now $\ln2$ (i.e. nonzero). 
So, the classical 
content of correlations can be increased by 
performing local general measurements and classically
communicating.

An important result was proved for pairs of spin--$1/2$ 
systems in \cite{Horodecki}: 
All states that are not of the form $\rho_{AB} = \sum_i p_i 
\rho^i_{A} \otimes \rho^i_{B}$, where $\sum_i p_i =1$ and 
$p_i \ge 0$ for all $i$, can be distilled to a subensemble of 
maximally entangled states using only operations $1$, $2$ 
and $3$. (The states of the above form obviously remain of 
the same form under any purification procedure). 
The local nature of the above three operations implies that 
we define a disentangled state of two 
quantum systems $A$ and $B$ as a state from which by means of 
local operations no subensemble of entangled states can be 
distilled. It should be noted that these states are sometimes
called separable in the existing literature. We also note
that it is not proven in general that if the state is not of this 
form then it can be purified.

\noindent
{\bf Definition 1}. A state $\rho_{AB}$ is disentangled iff
\begin{equation}
\rho_{AB} = \sum_i p_i \rho^i_{A} \otimes \rho^i_{B} \,\,\, ,
\label{sep}
\end{equation}
where, as before, $\sum_i p_i =1$ and $p_i \ge 0$ for all $i$. 
Otherwise it is said to be entangled. Note that all the states
in the above expansion can be taken to be pure. This is because each $\rho^i$
can be expanded in terms of its eigenvectors. So, in the above
sum we can in addition require that ${\rho_{A}^i}^2 =\rho_{A}^i$ and
${\rho_{B}^i}^2 = \rho_{B}^i$ for all $i$. This fact
will be used later in this section and will be formalized 
further in Section III.

\subsection{Quantification of Entanglement}

In the previous section we have indicated that out of certain 
states it is possible
to distill by means of LGM+CC+PS a subensemble of 
maximally entangled states (we call these states entangled). 
The question remains open about how much entanglement a 
certain state contains. Of course, this question is not
entirely well defined unless we state what physical
circumstances characterize the amount of entanglement.
This suggests that there is no unique measure
of entanglement. 
Before we define three different measures of
entanglement we state three conditions that every measure 
of entanglement has to satisfy. The third condition 
represents a generalization of the corresponding one in 
\cite{Vedral}.

\begin{description}
\item E1. $E({\sigma})=0$ iff ${\sigma}$ is separable.
\item E2. Local unitary operations leave $E({\sigma})$ 
invariant, i.e.
$E({\sigma})=E(U_A\otimes U_B \sigma
U_A^{\dagger}\otimes U_B^{\dagger})$.
\item E3. The expected entanglement cannot 
increase under 
LGM+CC+PS given by $\sum  V^{\dagger}_i V_i = {{1 \hspace{-0.3ex} \rule{0.1ex}{1.52ex}\rule[-.01ex]{0.3ex}{0.1ex}}}$, i.e. 
\begin{equation}
\sum tr(\sigma_i) \,\, E( \sigma_i/tr(\sigma_i)) 
\le E(\sigma)\;\; , 
\end{equation}
where $\sigma_i = V_i \sigma V^{\dagger}_i$.
\end{description}
Condition E1 ensures that disentangled and only 
disentangled states have a 
zero value of entanglement. Condition E2 ensures 
that a local change of basis has no effect on the 
amount of entanglement. Condition E3 is intended to 
remove the possibility of increasing entanglement
by performing local measurements aided by classical 
communication.
It is an improvement over the condition 3 
in \cite{Vedral} which required that  
$E(\sum_i V_i{\sigma}V^{\dagger}_i)\le E({\sigma})$. 
This condition E3 is physically more appropriate than that 
in \cite{Vedral} as it takes into account
the fact that we have some knowledge of 
the final state. Namely,
when we start with $n$ systems all in the state $\sigma$ we know 
exactly which $m_i = n\times tr(\sigma_i)$ pairs 
will end up in the state $\sigma_i$ after performing 
a purification procedure.  Therefore 
we can separately access the entanglement in 
each of the possible subensembles 
described by $\sigma_i$. Clearly the total 
expected entanglement at the end should not exceed the 
original entanglement, which is stated in E3. 
This, of course, does not 
exclude the possibility that we can select a 
subensemble whose entanglement per pair is
higher than the original entanglement per pair. 
We emphasise that if we assume that $E(\sigma)$ is also convex
(as it, indeed, is in the case of the Quantum Relative Entropy 
presented later in the paper) than E3 immediatelly implies that $E(\sum_i V_i{\sigma}V^{\dagger}_i)\le E({\sigma})$. On the other hand,
convexity of $E(\sigma)$ and $E(\sum_i V_i{\sigma}V^{\dagger}_i)\le E({\sigma})$
do not imply E3, which also provides a reason for requiring E3 rather 
than the condition in \cite{Vedral}. 
We now introduce three different measures of entanglement which 
obey E1--E3.

First we discuss the entanglement of creation \cite{Bennett1}. 
Bennett et al \cite{Bennett1} define the 
entanglement of creation of a state 
${\rho}$ by 
\begin{equation}
E_c(\rho):= \mbox{min} \sum_i p_i S(\rho_A^i)
\end{equation}
where $S(\rho_A)= -\mbox{tr} \rho_A \ln \rho_A$ is the von 
Neumann entropy and the minimum is taken over all the 
possible realisations of the state, 
$\rho_{AB} = \sum_j p_j |{\psi_j}\rangle\langle{\psi_j}|$ 
with $\rho_A^i =
\mbox{tr}_B (|{\psi_i}\rangle\langle{\psi_i}|)$. The
entanglement of creation satisfies 
all the three conditions E1--E3 \cite{Bennett1}. The
physical basis of this measure presents
the number of singlets needed to be shared in order to
create a given entangled state by local operations.  We will 
discuss this
in greater detail in Section IV. It should also be
added that progress has been made recently in finding a 
closed form of the entanglement of creation \cite{Wootters}.

Related to this measure is the entanglement of distillation
\cite{Bennett1}. It defines the amount of entanglement
of a state $\sigma$ as the proportion of singlets that 
can be distilled using a purification procedure (Bennett et al
distinguish one and two way communication which give rise to
two different measures, but we will not go into that much detail;
we assume the most general two way communication). As
such, it is dependent on the efficiency of a particular
purification procedure and can be made more general only by
introducing some sort of universal purification
procedure or asking for the best state dependent purification 
procedure. We investigate this in Section V.  
We now introduce our suggestion for a measure of 
an amount of entanglement.
It is seen in Section V that this measure is intimately
related to the entanglement of distillation by providing an
upper bound for it. 

If ${\cal D}$ is the set of all disentangled states, 
the measure of entanglement for a state $\sigma$ is 
then defined as

\begin{equation}
E({\sigma}):= \min_{\rho \in \cal D}\,\,\,
D(\sigma || \rho)
\label{measure}
	\label{6a}
\end{equation}  
where $D$ is any measure of {\em distance} 
(not necessarily a metric) between 
the two density matrices 
$\rho$ and $\sigma$ such that $E({\sigma})$ 
satisfies the above three conditions E1--E3 (see Fig. 1). 

Now the central question is what condition a candidate 
for $D(\sigma || \rho)$ has to satisfy in order 
for E1--E3 to hold for the entanglement measure? 
We present here a set of sufficient conditions.

\begin{description}
\item F1. $D(\sigma || \rho) \ge 0$ with the equality saturated 
iff ${\sigma}=\rho$.
\item F2. Unitary operations leave $D(\sigma||\rho)$ 
invariant, i.e.
$D(\sigma||\rho)=D(U\sigma U^{\dagger}||U \rho
U^{\dagger})$.
\item F3. $D(tr_p \sigma ||tr_p \rho) \le D({\sigma}||\rho)$, 
where $tr_p$ is a 
partial trace.
\item F4. $\sum p_i\, D(\sigma_i/p_i||\rho_i/q_i) \le
\sum D(\sigma_i||\rho_i)$, where $p_i = tr(\sigma_i)$, 
$q_i = tr(\rho_i)$ 
and $\sigma_i = V_i \sigma V^{\dagger}_i$ and 
$\rho_i = V_i \rho V^{\dagger}_i$ (note that $V_i$'s are
not necessarily local).
\item F5a. $D(\sum_i P_i\sigma P_i || 
\sum_i P_i \rho P_i) =\sum_i 
D(P_i\sigma P_i||P_i\rho P_i)$, where $P_i$ is any set of orthogonal
projectors such that $P_iP_j=\delta_{ij}P_i$.
\item F5b. $D(\sigma\otimes P_{\alpha}||\rho \otimes 
P_{\alpha})=D(\sigma||\rho) $
where $P_{\alpha}$ is any projector. 
\end{description}
 
Conditions F1 and F2 ensure that E1 and E2 hold;
F2, F3, F4 and F5 ensure that E3 is satisfied. The 
argument for the former is trivial, while for the 
latter it is more lengthy and will be presented 
in the remainder of this section.

\subsection{Proofs}

We claim that F2, F3, F4 and F5 are sufficient for E3 to be 
satisfied and hence need to prove that $F2 - F5 \Rightarrow E3$. 
If F2, F3 and F5b hold, then we can prove the following statement,

\noindent
{\bf Theorem 1}.  For any completely positive, trace 
preserving map $\Phi$, given by $\Phi \sigma =
\sum V_i\sigma V^{\dagger}_i$ and $\sum  V^{\dagger}_i 
V_i = {{1 \hspace{-0.3ex} \rule{0.1ex}{1.52ex}\rule[-.01ex]{0.3ex}{0.1ex}}}$, 
we have that  $D(\Phi \sigma ||\Phi \rho) 
\le D({\sigma}||\rho)$. \footnote{We frequently
interchange the $\Phi$ and $\sum V^{\dagger}V$ notations
for one another throughout this section.}

\noindent
{\bf Proof}. It is well known that 
a complete measurement can {\em always} be 
represented as a unitary operation+partial 
tracing on an extended Hilbert Space ${\cal H} 
\otimes {\cal H}_n$, where 
$dim {\cal H}_n = n$ \cite{Lindblad1,Lindblad2}. 
Let $\{|i\rangle \}$ be an 
orthonormal basis in  ${\cal H}_n$ and 
$|\alpha\rangle$ be a unit vector. So we define,                     
\begin{equation}
W= \sum_i V_i \otimes |i\rangle\langle \alpha | \,\, .
\label{a}
\end{equation}
Then, $W^{\dagger}W = {{1 \hspace{-0.3ex} \rule{0.1ex}{1.52ex}\rule[-.01ex]{0.3ex}{0.1ex}}}\otimes P_{\alpha}$ where 
$ P_{\alpha}= |\alpha 
\rangle\langle \alpha |$, and there is a unitary 
operator $U$ in ${\cal H} \otimes {\cal H}_n$ 
such that $W=U ({{1 \hspace{-0.3ex} \rule{0.1ex}{1.52ex}\rule[-.01ex]{0.3ex}{0.1ex}}}\otimes P_{\alpha})$ \cite{Reed}.
Consequently,
\begin{equation}  
U(A\otimes P_{\alpha})U^{\dagger} = \sum_{ij} 
V_iAV^{\dagger}_j \otimes 
|i\rangle\langle j|  \,\,\,\,   ,
\label{b}
\end {equation}
so that, 
\begin{equation}  
tr_2\{U(A\otimes P_{\alpha})U^{\dagger}\} = 
\sum_i V_iAV^{\dagger}_i  \,\,\,\, .
\end {equation} 
Now using F3, then F2, and finally F5b we find the following
\begin{eqnarray}
D(tr_2\{U(\sigma\otimes P_{\alpha})U^{\dagger}\}& 
|| & tr_2\{U(\rho\otimes 
P_{\alpha})U^{\dagger}\}) \\
& \le & D(U(\sigma\otimes P_{\alpha})U^{\dagger}
||U(\rho\otimes P_{\alpha})
U^{\dagger}) \\ 
& = & D(\sigma\otimes P_{\alpha}||\rho \otimes P_{\alpha}) \\
& = & D(\sigma||\rho) \,\,\,\,   .
\end{eqnarray}
This proves Theorem 1 ${}_{\Box}$.

\noindent
{\bf Corollary}. Since for a complete set of orthonormal 
projectors $P$, $\sum_i P_i \sigma P_i$ is 
a complete positive trace preserving map, then 
\begin{equation}
\sum_i D(P_i\sigma P_i||P_i \rho P_i) \le D(\sigma||\rho)\,\,\,\ .
\label{1}
\end{equation}
(The sum can be taken outside 
as F5a requires that $D(\sum_i P_i\sigma P_i || 
\sum_i P_i \rho P_i) =\sum_i 
D(P_i\sigma P_i||P_i\rho P_i)$). Now from F2, F3, F5b and 
eq.(\ref{1}) we have the following 

\noindent
{\bf Theorem 2}. If $\sigma_i = V_i \sigma 
V_i^{\dagger}$ then $\sum D(\sigma_i ||\rho_i) 
\le D({\sigma}||\rho)$.

\noindent
{\bf Proof}. Equations (\ref{a}) and (\ref{b}) are 
introduced as in the previous proof. 
From eq. (\ref{b}) we have that 
\begin{equation}  
tr_2\{{{\sf 1 \hspace{-0.3ex} \rule{0.1ex}{1.52ex}\rule[-.01ex]{0.3ex}{0.1ex}}} \otimes P_iU(A\otimes P_{\alpha})U^{\dagger} {{\sf 1 \hspace{-0.3ex} \rule{0.1ex}{1.52ex}\rule[-.01ex]{0.3ex}{0.1ex}}}
\otimes P_i\} = V_iAV^{\dagger}_i  \,\,\,\, .
\end {equation}
where $P_i = |i \rangle\langle i |$. Now, from F3, the 
Corollary and F5b it follows that 

\begin{eqnarray}
\sum_i & D & (tr_2\{{{\sf 1 \hspace{-0.3ex} \rule{0.1ex}{1.52ex}\rule[-.01ex]{0.3ex}{0.1ex}}} \otimes P_iU(\sigma\otimes P_{\alpha})
U^{\dagger}{{\sf 1 \hspace{-0.3ex} \rule{0.1ex}{1.52ex}\rule[-.01ex]{0.3ex}{0.1ex}}} \otimes P_i\}|| tr_2\{{{\sf 1 \hspace{-0.3ex} \rule{0.1ex}{1.52ex}\rule[-.01ex]{0.3ex}{0.1ex}}} \otimes P_i U(\rho\otimes 
P_{\alpha})U^{\dagger} {{\sf 1 \hspace{-0.3ex} \rule{0.1ex}{1.52ex}\rule[-.01ex]{0.3ex}{0.1ex}}} \otimes P_i\}) \\
& \le & \sum_i D({{\sf 1 \hspace{-0.3ex} \rule{0.1ex}{1.52ex}\rule[-.01ex]{0.3ex}{0.1ex}}} \otimes P_i U(\sigma\otimes P_{\alpha})
U^{\dagger}{{\sf 1 \hspace{-0.3ex} \rule{0.1ex}{1.52ex}\rule[-.01ex]{0.3ex}{0.1ex}}} \otimes P_i||{{\sf 1 \hspace{-0.3ex} \rule{0.1ex}{1.52ex}\rule[-.01ex]{0.3ex}{0.1ex}}} \otimes P_iU(\rho\otimes P_{\alpha})
U^{\dagger}{{\sf 1 \hspace{-0.3ex} \rule{0.1ex}{1.52ex}\rule[-.01ex]{0.3ex}{0.1ex}}} \otimes P_i) \\
& \le & D(U(\sigma\otimes P_{\alpha})U^{\dagger}||
U(\rho\otimes P_{\alpha})U^{\dagger})\\ 
& = & D(\sigma\otimes P_{\alpha}||\rho\otimes P_{\alpha}) \\
& = & D(\sigma||\rho) \,\,\,\,   .
\end{eqnarray}
This proves Theorem 2 ${}_{\Box}$.

\noindent
From Theorem 2 and F4 we have,
\begin{equation}
\sum p_i\, D(\frac{\sigma_i}{p_i}||\frac{\rho_i}{q_i}) \le
D(\sigma||\rho) \,\,\,\,\, .
\label{2}
\end{equation}
Now let $E(\sigma)=D(\sigma||\rho^*)$, {\em i.e.} 
let the minimum of $D(\sigma ||\rho)$ over 
all $\rho \in {\cal D}$ be attained at $\rho^*$. 
Then from eq. (\ref{2}) 
\begin{equation}
E(\sigma) := D(\sigma||\rho^*) \ge
\sum p_i\, D(\frac{\sigma_i}{p_i}||\frac{V^{\dagger}_i\rho^*V_i}{q_i}) \ge 
\sum p_i \,\, E(\sigma_i/p_i)
\label{ent}
\end{equation}
and E3 is satisfied. Note that in all the proofs 
for $D(\sigma||\rho)$ we 
never use the fact that the completely positive,
trace preserving map  
$\Phi$ is local. This is
only used in the last inequality of eq.(\ref{ent}) 
where LGM (+CC+PS)
maps disentangled states onto disentangled states. 
This ensures that $\rho^*_i$ is disentangled and 
therefore $D(\sigma_i/p_i||\rho^*_i/q_i) \ge E(\sigma_i/p_i)$.
So, the need for local $\Phi$ arises only in eq. (\ref{ent}); otherwise
all the other proofs hold for a general $\Phi$.
Note also that one can prove, by the same methods, a slightly more general
condition 
\begin{description}
\item $E3^{*}$. The expected entanglement of the initial state
$\sigma^n = \sigma_1 \otimes \ldots \otimes \sigma_n$ cannot increase under 
LGM+CC+PS given by $\sum  V^{\dagger}_i V_i = {{1 \hspace{-0.3ex} \rule{0.1ex}{1.52ex}\rule[-.01ex]{0.3ex}{0.1ex}}}$, i.e. 
\begin{equation}
	E(\sigma^n) \equiv E(\sigma_1 \otimes \ldots \otimes \sigma_n) \ge 
	\sum tr(V_i\sigma^n V_i^{\dagger}) 
	\,\, E( V_i\sigma^n V_i^{\dagger}/tr(V_i\sigma^n V_i^{\dagger})) \;\; .
\end{equation}
\end{description}
However, in the following we will not make use of this generalization.

\subsection{Two Realisations of $D(\sigma,\rho)$}

In this section we show that F1--F5 hold for the 
Quantum Relative Entropy and for the
Bures metric, which as we have seen immediately renders 
them generators of a good measure of entanglement.

\subsubsection{Quantum Relative Entropy}

We first prove F1--F5 for the Quantum Relative 
Entropy, i.e. when $D(\sigma||\rho)=S(\sigma||\rho):= 
\mbox{Tr}\, \left\{ \sigma (\ln  
\sigma - \ln \rho)\right\}$ (Note that the Quantum 
Relative Entropy is not a true metric, as it is not 
symmetric and does not satisfy the triangle inequality. 
In the next section the reasons for this will become clear. 
For further properties of the Quantum Relative Entropy
see \cite{Ohya,Petz,Donald2}.)
Properties F1 and F2 are satisfied \cite{Wehrl1}.
F3 follows from the strong subadditivity property of the 
von Neumann Entropy \cite{Lindblad1,Wehrl1,Lieb1,Lieb2}. Since 
$\sum S(\sigma_i||\rho_i) = \sum
p_i\, S(\sigma_i/p_i||\rho_i/q_i) + \sum 
p_i\ln p_i/q_i$ and $\sum p_i\ln\frac{p_i}{q_i} \ge 0$ 
(see \cite{Cover} for proof) F4 is also satisfied. 
Property F5 can be proved to hold by inspection \cite{Lindblad1}.
Now, a question arises as to why the entanglement is not 
defined as $E(\sigma) = \min_{\rho\in {\cal D}} 
S(\rho || \sigma)$. Since the Quantum Relative 
Entropy is asymmetric this gives a different result 
to the original definition. However, the major 
problem  with this convention is that for all pure states
this measure is infinite. Although this does have a 
sound statistical interpretation (see the next section) 
it is hard to relate it to any physically reasonable 
scheme (e.g. a purification procedure) and, in addition, it fails
to distinguish between different entangled pure states. This is the 
prime reason for excluding this convention from 
any further considerations. The measure 
of entanglement generated by the Quantum Relative Entropy 
will hereafter be referred to as the Relative Entropy of Entanglement.

\vspace*{0.5cm}
\noindent
{\em Properties of the Relative Entropy of Entanglement}

\vspace*{0.5cm}
\noindent
For pure, maximally entangled states we showed that 
the Relative Entropy of Entanglement reduces to the von Neumann 
reduced entropy \cite{Vedral}. We also conjectured 
\cite{Vedral} that for general pure state this 
would be true. Now we present a proof of this 
conjecture. In short, our proof goes as follows:
we already have a guess as to what the minimum for
a pure state $\sigma$ should be--say, it is a disentangled 
state $\rho^{*}$. Then we show that the gradient 
$\frac{d}{dx}S(\sigma || (1-x)\rho^{*} + x\rho)$ for any
$\rho \in {\cal D}$ is nonnegative. However, if $\rho^{*}$ 
was not a minimum the above gradient would be strictly
negative which is a contradiction. Now we present a more
formal proof \cite{Donpriv} that applies to arbitrary dimensions
of the two subsystems. An alternative proof that also applies to 
arbitrary dimensions will be given in section III. In the appendix 
we present a third proof that is restricted to two spin 1/2 systems but 
which can be generalized to arbitrary dimensions.

\noindent
{\bf Theorem 3.} For pure states $\sigma = \sum_{n_1n_2} \sqrt{p_{n_1}p_{n_2}} |\phi_{n_1}\psi_{n_1}\rangle\langle \phi_{n_2}\psi_{n_2} |$
the Relative Entropy of Entanglement
is equal to the Von Neumann reduced entropy, i.e. $E(\sigma) = -\sum_n
p_n \ln p_n$.

\noindent
{\bf Proof}.
For $a > 0$, $ \log a = \int_0^\infty {at - 1\over a + t} {dt \over
1 + t^2}$, and thus, for any positive operator $A$,
$ \log A = \int_0^\infty {At - 1\over A + t} {dt \over
1 + t^2}$. 
Let $f(x, \rho) = S(\sigma||(1-x) \rho^{*} + x \rho)$.  
Then
\begin{eqnarray}
 {\partial f \over \partial x}(0, \rho) & = & -\lim_{x \rightarrow 0}
\mbox{tr}\{{\sigma (\log ((1-x) \rho^{*} + x \rho) -
\log \rho^{*})\over x}\} \nonumber \\
&  =  & \mbox{tr} ( \sigma \int_0^\infty (\rho^{*} +
t)^{-1} ( \rho^{*} - \rho) (\rho^{*} +t)^{-1} dt \big) \nonumber  \\
& = & 1 -\int_0^\infty \mbox{tr}( \sigma (\rho^{*} + t)^{-1}  
\rho  (\rho^{*} +t)^{-1} \big) dt \nonumber \\
& = &
1 - \int_0^\infty \mbox{tr}( (\rho^{*} + t)^{-1} \sigma (\rho^{*} + t)^{-1}  
\rho \big) dt  
\end{eqnarray}

Take $\rho^{*} = \sum_n p_n |\phi_n\psi_n\rangle\langle \phi_n\psi_n|$ 
(this is our guess for the minimum).  Then
\begin{eqnarray}
(\rho^{*} + t)^{-1} \sigma (\rho^{*} + t)^{-1} & = & \sum_{n_1, n_2, n_3,
n_4} (p_{n_1} + t)^{-1}
|\phi_{n_1}\psi_{n_1}\rangle\langle\phi_{n_1}\psi_{n_1}| \nonumber \\
&   & 
\sqrt{p_{n_2}
p_{n_3}} |\phi_{n_2}\psi_{n_2}\rangle\langle\phi_{n_3}\psi_{n_3}| (p_{n_4} +
t)^{-1} |\phi_{n_4}\psi_{n_4}\rangle\langle\phi_{n_4}\psi_{n_4}| \nonumber \\
& = &
\sum_{n, n' } (p_{n} + t)^{-1} \sqrt{p_{n} p_{n'}}  (p_{n'} + t)^{-1} 
|\phi_{n}\psi_{n}\rangle\langle\phi_{n'}\psi_{n'}|. 
\end{eqnarray}

Set $g(p, q) = \int_0^\infty (p + t)^{-1} \sqrt{p q}  (q + t)^{-1} 
dt$. Then it follows that $g(p, p) = 1$ and, for $p < q$, 
\begin{eqnarray}
g(p, q) & = & \sqrt{p q}\int_0^\infty \big({1 \over p + t} - {1 \over q + 
t}\big) {1\over q - p} dt \\
& = & {\sqrt{p q}\over q - p} \log {q \over p} \;\; .
\end{eqnarray} 

\noindent
{\bf lemma}{\sl} $0 \leq g(p, q) \leq 1$ for all $p, q \in [0, 1]$.

\noindent
{\bf proof}. We know that $g(p, q) =  \sqrt{p q} \int_0^\infty (p + t)^{-1} 
(q + t)^{-1} dt$.
But,
\begin{equation}
(p + t)(q + t) = pq + t(p + q) + t^2 \geq pq + 2t\sqrt{p q} + t^2 = 
(\sqrt{pq} +t)^2 \;\; , 
\end{equation}
and so
\begin{equation}
g(p, q) \leq  \sqrt{p q} \int_0^\infty (\sqrt{pq} +t)^{-2} dt = 1\;\; .
\end{equation}

\noindent
Let $\rho = |\alpha\rangle\langle\alpha| \otimes |\beta\rangle\langle\beta|$ 
where $|\alpha\rangle =
\sum_n a_n |\phi_n\rangle$ and 
$\beta = \sum_n b_n \psi_n$ are normalized vectors.  Then
\begin{eqnarray}
{\partial f \over \partial x}(0, \rho) - 1 & = & - \mbox{tr}( \int_0^\infty 
(\rho^{*} + t)^{-1} \sigma (\rho^{*} + t)^{-1} dt \rho \big) \nonumber
\\
& = &  -\mbox{tr}(\sum_{n_1, n_2, n_3, n_4, n_5, n_6} g(p_{n_1}, p_{n_2}) 
|\phi_{n_1}\psi_{n_1}\rangle\langle\phi_{n_2}\psi_{n_2}| \nonumber \\
&   &
a_{n_3}
b_{n_4} \bar a_{n_5} \bar 
b_{n_6}|\phi_{n_3}\psi_{n_4}\rangle\langle\phi_{n_5}\psi_{n_6}|\big) 
\nonumber \\
& = &
-\sum_{n_1, n_2} g(p_{n_1}, p_{n_2})  a_{n_2} b_{n_2} \bar a_{n_1} \bar 
b_{n_1}  
\end{eqnarray}
and
\begin{eqnarray}
|{\partial f \over \partial x}(0, \rho) - 1|
& \leq & \sum_{n_1, n_2} |a_{n_1}| |b_{n_1}| |a_{n_2}| |b_{n_2}| \nonumber\\
& = & 
\big(\sum_{n} |a_{n}| |b_{n}|\big)^2 
\leq \sum_{n} |a_{n}|^2 \sum_n |b_{n}|^2 = 1 \;\; .
\end{eqnarray}
Thus it follows that ${\partial f \over \partial x}
(0, |\alpha\beta\rangle\langle\alpha\beta|)
\geq 0$.

But any $\rho \in {\cal D}$ can be written in the form $\rho = \sum_i r_i
|\alpha^i\beta^i\rangle\langle\alpha^i\beta^i|$ and so
${\partial f \over \partial x}(0, \rho) = \sum_i r_i {\partial f \over
\partial x}(0, |\alpha^i\beta^i\rangle\langle\alpha^i\beta^i|) \geq 0$.

\noindent
{\bf Proposition}{\sl} Let $\Phi \in H$ have Schmidt decomposition \cite{Schmidt} 
\begin{equation}
|\Phi\rangle = \sum_{n} \sqrt{p_n} |\varphi_n \psi_n\rangle
\end{equation}
and set $\sigma = |\Phi\rangle\langle\Phi|$. Then $E(\sigma) = -
\sum_n p_n \log p_n.$

\noindent
{\bf Proof}.  $S({\sigma} ||{\rho^{*}})  = -\sum_n p_n \log p_n$ so it is
sufficient to prove that $S({\sigma}||{\rho}) \geq
S({\sigma}||{\rho^{*}})$ for all $\rho \in
D$. Suppose that $S({\sigma}||{\rho}) <
S({\sigma}||{\rho^{*}})$ for some $\rho
\in {\cal D}$.  Then, for $0 < x \leq 1$,
\begin{eqnarray}
 f(x, \rho) = S({\sigma}||{(1-x) \rho^{*} + x \rho}) & \leq & (1-x)
S({\sigma}||{\rho^{*}}) + x S({\sigma}||{\rho}) \nonumber \\
& = & (1-x) f(0,\rho) + x f(1, \rho) \;\; .
\end{eqnarray}
This implies
\begin{equation}
{f(x, \rho) - f(0, \rho) \over x} \leq f(1, \rho) - f(0, 
\rho) < 0 \;\; .  
\end{equation}
This is impossible since 
${\partial f \over \partial x}(0, \rho) = \lim_{x \rightarrow 0} {f(x, 
\rho) - f(0, \rho) \over x} \geq 0$.  This therefore proves the
above proposition ${}_{\Box}$. 

Therefore we have shown that for arbitrary dimensions of the subsystems
the entropy of entanglement reduces to the entropy of entanglement for pure 
states. This is, in fact, a very desirable property, as the entropy of entanglement
is known to be a good measure of entanglement for pure states. In fact 
one might want to elevate Theorem 3 to a condition for any good measure 
of entanglement, i.e.
\begin{description}
\item E4: For pure states the measure of entanglement reduces to the entropy of
entanglement, i.e.
\begin{equation}
	E(\sigma) = - tr \left\{ \sigma_A \ln \sigma_A \right\} \;\; ,
\end{equation}
with $\sigma_A=tr_B\{\sigma\}$ being the reduced density operator of one subsystem
of the entangled pair.
\end{description}
However, in subsection 2 we will see that measures which do not satisfy 
E4 can nevertheless contain useful information. We will discuss this point 
later in this paper.

We would like to point out another property of the relative entropy of
entanglement that helps us find the amount of entanglement. It gives us 
a method to construct from a density operator $\sigma$ with known entanglement
a new density operator $\sigma'$ with known entanglement. 

\noindent
{\bf Theorem 4}. 
If $\rho^*$ minimizes $S(\sigma ||\rho^*)$ over $\rho \in {\cal D}$  then $\rho^*$ is
also a minimum for any state of the form
$\sigma_x= (1-x)\sigma + x\rho^*$.

\noindent
{\bf Proof}. Consider,
\begin{eqnarray}
S(\sigma_x ||\rho) - S(\sigma_x ||\rho^*) & = & \mbox{tr} \{\sigma_x \ln
\rho^* - \sigma_x \ln \rho \} \nonumber \\
& = & -x \mbox{tr} (\sigma\ln \rho) - (1-x)\mbox{tr}(\rho^*\ln \rho)
+ x\mbox{tr} (\sigma\ln \rho^*) + (1-x)\mbox{tr}(\rho^*\ln \rho^*) \nonumber \\
& = & x \{ S(\sigma ||\rho)-S(\sigma || \rho^*)\} + (1-x) S(\rho^* ||\rho) \ge 0
\end{eqnarray} 
This is true for any $\rho$. Thus $\rho^*$ is indeed a minimum of $\sigma_x$
${}_{\Box}$. For completeness we now prove here that $E(\sigma)$ is convex.
Namely, 

\noindent
{\bf Theorem 5.} 
$E(x_1 \sigma_1 + x_2 \sigma_2) \le x_1 E(\sigma_1) + x_2 E(\sigma_2)$,
where $x_1 + x_2 =1$.

\noindent
{\bf Proof}. This property follows from the convexity of the
quantum relative entropy in both arguments \cite{Donald2}
\begin{equation}
S(x_1 \sigma_1 + x_2 \sigma_2 ||x_1 \rho_1 + x_2 \rho_2 ) \le x_1 
S(\sigma_1 ||\rho_1) + x_2 S(\sigma_2 ||\rho_2) \;\;.
\end{equation}
Now, 
\begin{eqnarray}
E(x_1 \sigma_1 + x_2 \sigma_2) & \le & S(x_1 \sigma_1 + x_2 \sigma_2 || 
x_1\rho_1^*+ x_2 \rho_2^*) \nonumber \\
& \le & x_1 S(\sigma_1||\rho_1^*) + x_2 S(\sigma_2 ||\rho_2^*) \nonumber \\
& = & x_1 E(\sigma_1) + x_2 E(\sigma_2) \;\; ,
\end{eqnarray}
which completes our proof of convexity ${}_{\Box}$. This is physically
a very satisfying property of an entanglement measure. It says that 
when we mix two states having a certain amount of
entanglement we cannot get a more entangled state, i.e.
succinctly stated ``mixing does not increase entanglement".
This is what is indeed expected from a measure of entanglement
to predict. 

As a last property we state that the entanglement of creation $E_c$ 
is never smaller 
than the Relative Entropy of Entanglement $E$. We will show later
that this property has the important implication that the amount of entanglement
that we have to invest to create a given quantum state is usually larger
than the entanglement that you can recover using quantum state distillation
methods.
   
\noindent
{\bf Theorem 6.} $E(\sigma) \le E_c(\sigma) = \min_{\rho \in {\cal D}} 
S(\sigma ||\rho)$.

\noindent
{\bf Proof}. Given a state $\sigma$ then by definition of the entanglement 
of creation there is a convex decomposition $\sigma=\sum p_i \sigma_i$ with 
pure states $\sigma_i$ such that
\begin{equation}
	E_c(\sigma) = \sum p_i E_c(\sigma_i)\;\; .
\end{equation}
As the entanglement of creation coincides with our entanglement for pure 
states and as 
our entanglement is convex it follows that
\begin{equation}
E_c(\sigma) = \sum p_i E_c(\sigma_i) = \sum p_i E(\sigma_i) \ge 
E(\sum p_i\sigma_i)=E(\sigma)\;\; ,
\end{equation}
and the proof is completed ${}_{\Box}$.

The physical explination of the above result lies in the fact that
a certain amount of additional knowledge is involved in the
entanglement of formation which gives it a higher value to the
Relative Entropy of Entanglement. This will be explained in full 
detail in section V.
We add that the relative entropy of entanglement $E(\sigma)$ can be 
calculated easily for Bell diagonal states \cite{Vedral}. Comparing the 
result to those for the entanglement of creation \cite{Bennett1}
one finds that, in fact, strict inequality holds. 
In general, we have 
unfortunately found no ``closed form" for the relative entropy of
entanglement and a computer search is 
necessary to find the minimum $\rho^*$, for each given
$\sigma$. However, we can numerically find the amount
of entanglement for two spin $1/2$ subsystems very efficiently 
using general methods independent of the dimensionality and
the number of subsystems involved which are described in the next section.

\subsubsection{Bures Metric}

Another distance measure that leads to a measure of entanglement that
satisfies the conditions E1-E3 is induced by the Bures metric. However,
it will turn out that it does not satisfy condition E4 and is therefore
a less useful measure. In fact some people would say it is not a measure
of entanglement at all, however, we believe that this very much depends
on the questions one asks.

We now prove F1-F5 for the Bures metric, i.e. when 
$D(\sigma||\rho)=D_B(\sigma||\rho) := 2- 2\sqrt{F(\sigma,\rho)}$, 
where $F(\sigma,\rho):= \left[tr\{\sqrt{\rho}\sigma
\sqrt{\rho}\}^{1/2}\right]^2$ is the so called 
fidelity (or Uhlmann's transition probability). Property F1 follows 
from the fact that the Bures metric is a true metric and F2 is obvious. 
F3 is a consequence of the fact that $D_B$ does not increase under a 
complete positive trace-preserving map \cite{Barnum}. We can also 
easily check that $p_iq_i F(\sigma_i/p_i,\rho_i/q_i)= F(\sigma_i,\rho_i)$, 
from where F4 immediately follows as $q_i \in [0,1]$. F5 is seen
to be true by inspection. 
As conditions F1-F5 are satisfied, it immediately follows that
conditions E1-E3 are satisfied too. 

In the following present some properties of the
Bures measure of entanglement $E_B(\sigma)$.
First we show that for pure states we do not recover
the entropy of entanglement.

\noindent
{\bf Theorem 7:} For a pure state 
$|\psi\rangle = \alpha |00\rangle + \beta |11\rangle$ one has
\begin{equation}
	E_B(|\psi\rangle \langle \psi |) = 4 \alpha^2 (1 - \alpha^2)\;\; .
	\label{theorem7}
\end{equation}
{\bf Proof.} To prove Theorem 7 we have to show that the closest disentangled 
state to $\sigma=|\psi\rangle \langle \psi |$ under the Bures metric is given 
by $\rho^{*}= \alpha^2 |00\rangle\langle 00| + \beta^2 |11\rangle\langle 11|$
To this end we consider a slight variation around $\rho^{*}$ of the form
$\rho_{\lambda} = (1-\lambda) \rho^{*} + \lambda \rho$ where $\rho\in {\cal D}$.
Now we need to calculate 
\begin{equation}
	\frac{d}{d\lambda} D_B(\sigma||\rho_{\lambda}) |_{\lambda=0}
	= \frac{d}{d\lambda} tr\left\{\sqrt{\sqrt{\sigma}\rho_{\lambda}\sqrt{\sigma}}
	\right\} \le 0 
\end{equation}
Using the fact that $\sqrt{\sigma}=\sigma$ as $\sigma$ is pure we obtain
\begin{eqnarray}
	\frac{d}{d\lambda} D_B(\sigma||\rho_{\lambda}) |_{\lambda=0} =
	 \frac{d}{d\lambda} \sqrt{ \alpha^4 +\beta^4 + 
	\lambda ( \langle \psi| \rho |\psi \rangle  - 1)} |_{\lambda=0} \le 0 \;\; .
\end{eqnarray}
Using the closest state $\rho^{*}$ one then obtains eq.(\ref{theorem7}). To obtain
the entanglement of an arbitrary pure state one first has to calculate the 
Schmidt decomposition \cite{Schmidt} and than by local unitary transformation 
transform the
state to the form $|\psi\rangle = \alpha |00\rangle + \beta |11\rangle$. As local
unitary transformations do not change the entanglement, we have therefore
shown that the Bures measure of entanglement does not reduce to the entropy of 
entanglement for pure states.
The proof presented here can be generalized to many dimensional systems but we
do not state this generalization.

In fact, it is now easy to see the following 

\noindent
{\bf Corollary.} The Bures measure of entanglement for pure
states is smaller than the entropy of entanglement, i.e. for any pure state $\sigma$
\begin{equation}
	E_B(\sigma) \le - tr \left\{ \sigma_A \ln \sigma_A \right\}
	\label{Corrolary1}
\end{equation}
{\bf Proof.} One can see quickly that for $\alpha\in [0,1]$
\begin{equation}
	4 \alpha^2 (1-\alpha^2) \le -\alpha^2 \ln \alpha^2 - 
	(1-\alpha^2) \ln (1-\alpha^2)
\end{equation}
from which the Corollary follows.

As the Bures measure of entanglement does not satisfy condition E4, i.e. does
not reduce to the entropy of entanglement for pure states one might argue
that it does not provide a sensible measure of entanglement. However, it should
be noted that the Bures metric immediately gives an upper bound on the following
very special purification procedure. Assume that Alice and Bob 
are given EPR pairs, but one pair at
a time. Then they are allowed to perform any local operations they like, and then 
decide whether we keep the pair or discard it. Then, they are given the next EPR pair.
The question is, how many pure singlet states they can possibly distill out of 
such a purification procedure. The answer is immediately obvious from 
condition E3. The best that Alice and Bob can do is to have one 
subensemble with pure singlets
and all other subensembles with disentangled states. Then the probability to
obtain a singlet is simply given by the Bures measure of entanglement 
for the initial ensemble. As this is smaller than the entropy of entanglement 
we have found the nontrivial, though not very surprising result, that this
restricted purification procedure is strictly less efficient than 
entanglement concentration described in \cite{Bennett}. 

\subsubsection{Other candidates}

A reasonable candidate to generate a measure of entanglement 
is the Hilbert--Schmidt Metric. Here we have that 
$D(A||B)=||A-B||^2:=tr (A-B)^2$. 
F1 follows from the fact that $||A-B||$ is a 
true metric, and F2 is obvious. 
F3 and F4 remain to be shown to hold. 
We also believe that there are numerous
other nontrivial choices for $D(A||B)$ 
(by nontrivial we mean that the choice is not a simple scale 
transformation of the above candidates).
Each of those generators would arise from a different physical
procedure involving measurements conducted on $\sigma$ and $\rho^*$.
None of the choices could be said to be
more important than any other {\em a priori}, but the significance
of each generator would have to be seen through 
physical assumptions. To illustrate this point further, let us 
take an extreme example. Define,
\[
D(A || B) = \left \{ \begin{array}{r@{\quad:\quad}l}
			1 & A\not= B \\ 0 & A=B
				\end{array} \right . \;\; .  
\]   
If entanglement is calculated using this distance, then 
\[
E(\sigma) = \left\{ \begin{array}{r@{\quad:\quad}l}
	1 & \sigma \notin {\cal D}  \\ 0 & \sigma \in {\cal} D
		\end{array} \right.  \;\; . 
\] 
This measure therefore tells us if a given state $\sigma$ is entangled, 
i.e. when $E(\sigma) = 1$, or disentangled, i.e. when $E(\sigma) = 0 $. 
We can call it the ``indicator measure" of entanglement. It should be noted 
that this measure trivially satisfies conditions E1-E3. This shows
that there are numerous different choices for $D(A || B)$ and 
each is related to different physical considerations.  
We explain the statistical basis of the 
Relative Entropy of Entanglement in Section IV.
The Relative Entropy of Entanglement is then seen to be linked
very naturally to the notion of a purification procedure.
First, however, we present an efficient numerical method to 
obtain entanglement for arbitrary particles.

\section{Numerics for Two Spin $1/2$ Particles}

In order to understand how our program for calculating the amount
of entanglement works, we first need to introduce one basic definition
and one important result from convex analysis \cite{Rock}. From this
point onwards we concentrate on the Quantum Relative Entropy as a
measure of entanglement although most of the considerations
are of a more general nature. 

\noindent
{\bf Definition 2}. The convex hull ($\mbox{co}(A)$) of a set $A$ is the
set of all points which can be expressed as (finite) convex combinations of
points in $A$.  In other words, $x \in \mbox{co}(A)$ if and only if
$x$ has an expression of the form $x = \sum_{k=1}^{K} p_k a_k$ where $K$
is finite, $\sum_{k=1}^{K} p_k = 1$, and, for $k = 1, \dots, K$, $p_k > 
0$ and $a_k \in A$.

We immediately see that the set of disentangled states ${\cal D}$ is a 
convex hull of its pure states. This means that any state in ${\cal D}$
can be written as a convex combination of the form 
$\sum p_n |\phi_n\psi_n\rangle\langle \phi_n\psi_n |$. However, there is
now a problem in the numerical determination of the measure of entanglement.
We have to perform a search over the set of disentangled states in order 
to find that disentangled state which is closest to the state $\sigma$ of 
which we want to know the entanglement. But how can we parametrize the 
disentangled 
states? We know that the disentangled states are of the form given by 
Definition 1. However, there the number of states in the convex combination 
is not limited. Therefore one could think that we have to look over all 
convex combinations with one state, then two states, then 1000 states and 
so forth. The next theorem, however, shows that one can put an upper limit 
to the number of states that are required in the convex combination. This 
is crucial for our minimization problem as it shows that we do not have to 
have an infinite number of parameters to search over.

\noindent
{\bf Caratheodory's theorem}.  Let $A \subset {\bf R}^N$.  Then any
$x \in \mbox{co}(A)$ has an expression of the form
$x = \sum_{n=1}^{N+1} p_n a_n$ where $\sum_{n=1}^{N+1} p_n = 1$, and, for
$n = 1, \dots, N+1$, $p_n \geq 0$ and $a_n \in A$.

A direct consequence of Caratheodory's theorem is that 
any state in ${\cal D}$ can be decomposed into a sum of at most 
$(\mbox{dim}(H_1) \times \mbox{dim}(H_2))^2$ 
products of pure states. So, for 2 spin$1/2$ particles there
are at most $16$ terms in the expansion of any disentangled
state. In addition, each pure state can be described using 
two real numbers, so that there are altogether at most
$15 + 16\times 4 = 79$ real parameters needed to completely
characterize a disentangled state in this case. 

A random search over the $79$ real parameters would still be very 
inefficient. However, we can now make use of another useful property 
of the relative entropy, which is the fact that it is convex. 
This means that we have to 
minimize a convex function over the convex set of disentangled states.
It can easily be shown that any local minimum must 
also be a global minimum. Therefore we can perform a gradient search
for the minimum (basically we calculate the gradient and then perform a 
step in the opposite direction and repeat this procedure until 
we hit the minimum). As  soon as we have found any 
relative minimum we can stop the 
search, since this is also a global minimum. To make the 
gradient search efficient
we have to chose a suitable parametrization. The 
parametrization that we use has the advantage that
it also provides us with another proof of Theorem 3 which states that
for pure states the Relative Entropy of Entanglement reduces to the
von Neumann reduced entropy. We first explain the parametrization 
and then state the alternative proof for Theorem 3. 
The following results can easily be extended to two subsystems of arbitrary
dimensions but for clarity we restrict ourselves to two spin $1/2$
systems.

Our aim is to find the amount of entanglement of a state 
$\sigma$ of two spin 1/2 states, i.e. we have to minimize 
$tr\{\sigma \ln\sigma - \sigma\ln\rho\}$ 
for all $\rho\in{\cal D}$. From Caratheodory's theorem we know that we 
only need convex combinations of at most $16$ pure states $\rho_k^i$ to 
represent $\rho\in{\cal D}$, 
i.e.
\begin{equation}
	\rho = \sum_{i=1}^{16} p^2_i \rho_{1}^i \otimes \rho_{2}^i\;\; .
	\label{numerics1}
\end{equation}
(Notice that we use $p_i^2$ instead of $p_i$ for convenience, so that
here we require that $\sum_{i=1}^{16} p^2_i =1$).
The parametrization we chose is now given by
\begin{equation}
	p_i = \sin\phi_{i-1} \prod_{j=i}^{15} \cos\phi_j \;\; \mbox{with} \;\;
	\phi_0 = \frac{\pi}{2}
	\label{numerics2}
\end{equation}
and 
\begin{eqnarray}
	\rho_k^{i} &=& |\psi_k^i\rangle\langle \psi_k^i| \nonumber \\
	|\psi_1^i\rangle &=& \cos\alpha_i |0\rangle + 
			     \sin\alpha_i e^{i\,\eta_i} |1\rangle \label{numerics3}\\
	|\psi_2^i\rangle &=& \cos\beta_i |0\rangle + 
			     \sin\beta_i e^{i\,\mu_i} |1\rangle 
	\nonumber
\end{eqnarray}
All angles $\alpha_i,\beta_i,\phi_i,\eta_i,\mu_i$ can have arbitrary values, but 
due to the periodicity only the interval $[0,2\pi]$ is really relevant. 
Numerically this has the advantage that our parameter space has no edges 
at which problems might occur. The program for the search of the minimum 
is now quite straightforward. The idea is that given $\sigma$ 
we start from a random $\rho$, i.e. we generate $79$ 
random numbers. Then we compute $S(\sigma||\rho)$, as well as 
small variations of the $79$ parameters of $\rho$, 
to obtain the approximate gradient of $S(\sigma||\rho)$ at the point $\rho$.
We then move opposite to the gradient to
obtain the next $\rho$. We continue this until we reach the
minimum. As explained before, a convex function over a 
convex set can only have a global minimum, so that 
the minimum value we end up with is the
one and only. The method outlined above immediately generalizes to 
two subsystems of arbitrary dimension, however, the number of parameters
rises quickly to large values which slows down the program considerably.

Before we state some numerical results we now indicate an alternative 
proof of Theorem 3 using Caratheodory's theorem and the parametrization 
given in eqs. (\ref{numerics1}) - (\ref{numerics3}). For this proof we use
the fact that we can represent the logarithm of an operator $\rho$ by
\begin{equation}
	\ln\rho = \frac{1}{2\pi i} \oint \ln z \frac{1}{z{{\sf 1 \hspace{-0.3ex} \rule{0.1ex}{1.52ex}\rule[-.01ex]{0.3ex}{0.1ex}}} - \rho}
	\label{proofnew1}
\end{equation}
where the path of integration encloses all eigenvalues of $\rho$.
We can now take the partial derivative of $\ln\rho$ with respect to a
parameter $\phi$ on which $\rho$ might depend. 
\begin{equation}
	\frac{\partial\ln\rho}{\partial\phi} = 
	\frac{1}{2\pi i} \oint \ln z \frac{1}{z{{\sf 1 \hspace{-0.3ex} \rule{0.1ex}{1.52ex}\rule[-.01ex]{0.3ex}{0.1ex}}} - \rho}\;
	\frac{\partial\rho}{\partial\phi} \;\frac{1}{z{{\sf 1 \hspace{-0.3ex} \rule{0.1ex}{1.52ex}\rule[-.01ex]{0.3ex}{0.1ex}}} - \rho} \;\; .
	\label{proofnew2}
\end{equation}
Now, we have a given pure state 
\begin{equation}
	\sigma = \alpha^2 |00\rangle\langle 00|  + \alpha\sqrt{1-\alpha^2}
	(|00\rangle\langle 11| + |11\rangle\langle 00|) 
	+ (1-\alpha^2) |11\rangle\langle 11|
	\label{proofnew3}
\end{equation}
The suspected closest approximation to $\sigma$ within the disentangled states
is given by
\begin{equation}
	\rho_{min} = \alpha^2 |00\rangle\langle 00| +  
		 (1-\alpha^2) |11\rangle\langle 11| \;\; .
	\label{proofnew4}
\end{equation}
If we want to represent $\rho_{min}$ using the parametrization given in
eqs. (\ref{numerics1}) - (\ref{numerics3}) then we find for these parameters
$\cos^2\phi_1 = \alpha^2\; ; \; \alpha_2=\beta_2=\frac{\pi}{2}$ and zero for
all other parameters. Using eq. (\ref{proofnew2}) one can now calculate all 
the partial
derivatives of the relative entropy around the point $\rho_{min}$. It is 
easy, but rather lengthy, to check that these derivatives vanish and that 
therefore $\rho_{min}$ is a relative minimum. This concludes the proof as 
a relative minimum of a convex function on a convex set is also a global 
minimum.

After this additional proof of Theorem 3 we now state some results
that we have obtained or confirmed with the program that implements 
the gradient search. We present four nontrivial states $\sigma$ 
for which we can find the closest disentangled  
state $\rho$ that minimize the Quantum Relative Entropy
thereby giving the Relative Entropy of Entanglement. Using the same 
ideas as for the proof of Theorem 3 in Eq. (\ref{proofnew1} - \ref{proofnew4})
one can then prove that these are indeed the closest disentangled states.

\noindent
{\bf Example 1}.
\begin{eqnarray}
& \sigma_1 & =  \lambda |\Phi^+\rangle \langle \Phi^+| +
(1-\lambda) |01\rangle\langle 01| \\
& \rho_1 & =   {\lambda \over 2} (1-{\lambda \over 2})|00\rangle\langle 00| + 
{\lambda\over 2} (1-{\lambda \over 2})\{ |00\rangle\langle 11| +\mbox{H.C.}\} 
+ \nonumber \\
&        & (1- {\lambda \over 2})^2 |01\rangle\langle 01| +
{\lambda^2 \over 4} |10\rangle\langle 10| + {\lambda \over 2} 
(1-{\lambda\over 2})|11\rangle\langle 11| \\
& E(\sigma_1) & =  (\lambda -2)\ln (1-\frac{\lambda}{2}) + (1-\lambda)
\ln (1-\lambda)  \;\; .
\end{eqnarray}
Here $|\Phi^+\rangle$ is one of the four Bell states defined by
\begin{eqnarray}
|\Phi^{\pm}\rangle & = & \frac{1}{\sqrt{2}} (|00\rangle \pm |11\rangle ) \\
|\Psi^{\pm}\rangle & = & \frac{1}{\sqrt{2}} (|01\rangle \pm |10\rangle )
\end{eqnarray}

\noindent
{\bf Example 2}.
\begin{eqnarray}
& \sigma_2 & = \lambda |\Phi^+\rangle \langle \Phi^+| +(1-\lambda) |00
\rangle\langle 00| \\
& \rho_2 & =  (1-\frac{\lambda}{2}) |00\rangle\langle 00| + \frac{\lambda}{2} 
|11\rangle\langle 11|\\
& E(\sigma_2) & = s_+ \ln s_+ + s_- \ln s_- - (1-\frac{\lambda}{2}) \ln (1-\frac{\lambda}{2}) - (1-\frac{\lambda}{2})\ln (1-\frac{\lambda}{2}) \;\; ,
\end{eqnarray} 
where 
\begin{equation}
s_{\pm} = \frac{1\pm \sqrt{1-2\lambda (1-\frac{\lambda}{2})}}{2}
\end{equation}
are the eigenvalues of $\sigma_2$.
One could argue that in the above two cases the following reasoning
can be applied: $\sigma_{1(2)}$ 
is a mixture of a maximally entangled state (for which the amount of
entanglement is given by $\ln 2$) and a completely disentangled 
state ($E= 0$). Thus 
one would expect a total amount of entanglement of $\lambda \ln 2$. It is
curious that this reasoning does not work for either of the two states, 
since, in fact, $E(\sigma_{1(2)}) \le \lambda \ln 2$.   
Now, we show how to use Theorem 4 to generate more states and 
their minima. For pure states $\sigma^2 = \sigma $ we know the
minimum $\rho$. Now, the state that is a convex sum of $\sigma$ and
$\rho$ should also have the same minimum $\rho$. So,
 
\noindent 
{\bf Example 3}.
\begin{eqnarray}
& \sigma_3 & = A |00\rangle\langle 00| + B |00\rangle\langle 11| + 
B^* |11\rangle\langle 00| + 
(1-A) |11\rangle\langle 11| \\
& \rho_3 & = A |00\rangle\langle 00| + (1-A) |11\rangle\langle 11|  \\
& E(\sigma_3) & = e_+ \ln e_+ + e_- \ln e_- - A\ln A -(1-A) \ln (1-A) \;\; ,
\end{eqnarray}
where
\begin{equation}
e_{\pm} = \frac{1 \pm \sqrt{1-4A(1-A) - |B|^2}}{2} \;\; .
\end{equation}
Using Theorem 4, the amount of entanglement can be found for a number of
other spin $1/2$ states. Our program can also help us infer the 
entanglement of some other non-trivial states as the last example
shows.

\noindent 
{\bf Example 4}.
\begin{eqnarray}
& \sigma_4 & = A |00\rangle\langle 00| + B |00\rangle\langle 11| + 
B^* |11\rangle\langle 00| + (1-2A)|01\rangle\langle 01| +
A |11\rangle\langle 11| \\
& \rho_4 & = C |00\rangle\langle 00| + D |00\rangle\langle 11| + 
D^* |11\rangle\langle 00| + E|01\rangle\langle 01| \\
&   & + (1-2C-E)|10\rangle\langle 10| + C |11\rangle\langle 11|  \;\; ,
\end{eqnarray}
where 
\begin{eqnarray}
& E & = \frac{(1-2A)(1-A)^2}{(1-A)^2-B^2} \\
& C & = 1- A - E \\
& D & = \sqrt{E(1-E -2C)} = \frac{(1-2A)(1-A)}{(1-A)^2-B^2}B \;\; .
\end{eqnarray}
It is now easy to compute the amount of entanglement from the
above information.

In addition to the above described methods there is a simple way
of obtaining a lower bound for the amount of entanglement for
any two spin $1/2$ system. Suppose that we have a
certain state $\sigma$. We first find the {\em maximally}
entangled state $|\psi\rangle$ such that the fidelity 
$F=\langle \psi | \sigma |\psi\rangle$ is maximized. Then we
apply local unitary transformations to $\sigma$ which transform 
$|\psi\rangle$ into the singlet state (this is, of course, always
possible). Now, we apply local random rotations \cite{Bennett1} to
both particles. These will transform $\sigma$ into a Werner state,
where the singlet state will have a weight $F$  (since it is invariant under
rotations) and all the other three Bell states will have equal 
weights of $(1-F)/3$ (since they are randomized). 
Since these operations are local they cannot
increase the amount of entanglement, and we have that for any $\sigma$

\begin{equation}
E(\sigma) \ge E(W_F) = F\ln F +(1-F)\ln (1-F) + \ln 2
\end{equation}      
where $W_F$ is the above described Werner state (the Relative Entropy of entanglement for a general Bell diagonal state is calculated in \cite{Vedral}).

We note that this efficient computer 
search provides an alternative criterion for deciding when a given
state $\sigma $ of two spin $1/2$ systems is disentangled, i.e. 
of the form given in eq. (\ref{sep}). The already existing criterion is 
the one given by Peres and Horodecki family (see second and third references in \cite{Gisin}), which states that a
state is disentangled iff its partial trace over either of the subsystems
is a non-negative operator. This criterion
is only valid for two spin $1/2$, or one spin $1/2$ and one spin $1$
systems. In the absence of a more general analytical criterion our
computational method provides a way of deciding this question. In 
addition we would like to point out that the program is also able to
provide us with the convex decomposition of a disentangled state $\rho$.

At the end of this section we mention {\em additivity} as an important 
property desired from a measure of entanglement, i.e. we would like to 
have

\begin{equation}
E(\sigma_{12}\otimes\sigma_{34}) = E(\sigma_{12}) + E(\sigma_{34}) \,\,\, ,
\label{additive}
\end{equation} 
where systems $1+2$ and systems $3+4$ are
entangled separately from each other. The exact
definition of the left hand side is
\begin{equation}
E(\sigma_{12}\otimes\sigma_{34}) = \min_{p_i,\rho_{13},\rho_{24}} 
S(\sigma_{12}\otimes\sigma_{34} ||
\sum_i p_i \rho_{13}^i\otimes\rho_{24}^i)  \;\; .
\label{add}
\end{equation}
Why this form? One would originally assume that  
$\sigma_{12}\otimes\sigma_{34}$ should be minimized by the states 
of the form $(\sum_i p_i \rho_{1}^i\otimes 
\rho_{2}^i)\otimes(\sum_j p_j \rho_{3}^j\otimes \rho_{4}^j)$. However,
Alice, who holds systems $1$ and $3$, and Bob, who holds systems $2$ and $4$, 
can also perform arbitrary unitary operation on 
their subsystems (i.e. locally). This obviously leads to the creation of
entanglement between 1 and 3 and between 2 and 4 and hence the
form given in eq. (\ref{add}).  
Additivity is, of course, already true for the pure states,
as can be seen from the proof above, when our measure
reduces to the von Neumann entropy.
For more general cases we were unable to provide an
analytical proof, so that the above additivity property
remains a conjecture. However, for two spin $1/2$ systems,
our program did not find any counter-example.
It should be noted that it is easy to see that we have 
\begin{equation}
	E(\sigma_{12}\otimes\sigma_{34}) \le E(\sigma_{12}) + E(\sigma_{34}) \,\, .
\end{equation}
In the following we will assume that Eq. (\ref{additive}) holds
and use it in Section V to derive certain limits
to the efficiency of purification procedures.    

\section{Statistical Basis of Entanglement Measure}

Let us see how we can interpret our entanglement measure in 
the light of experiments, i.e. statistically. This was presented in 
\cite{Vedral2} in a greater detail. Here we present a summary
which is sufficient to understand the following Section. Our 
interpretation relies on the result concerning the asymptotics 
of the Quantum Relative Entropy first proved in \cite{Petz}, and
here presented under the name of Quantum Sanov's Theorem. 
We first show how the notion of Relative Entropy arises in 
classical information theory as a measure of distinguishability 
of two
probability distributions. We then generalize this idea to 
the quantum case, i.e. to distinguishing between two quantum 
states (for a discussion of distinguishability of pure
quantum states see e.g. 
\cite{Wootters1}). We will see
that this naturally leads to the notion of the Quantum 
Relative Entropy. It is then straightforward to extend this
concept to explain the Relative Entropy of Entanglement. 
Suppose we 
would like to check if a given coin is ``fair", i.e. if
it generates a ``head--tail" distribution of $f=(1/2,1/2)$. 
If the coin is biased then it will produce some other
distribution, say $uf=(1/3,2/3)$. So, our question of
the coin fairness boils down to how well we can differentiate
between two given probability distributions given a finite, n,
number of experiments to perform on one of the two distributions. 
In the case of a coin we would toss it
n times and record the number of 0's and 1's. From simple statistics
we know that if the coin is fair than the number of 0's $N(0)$ will
be roughly $n/2-\sqrt{n} \le N(0) \le n/2+\sqrt{n}$, for large $n$ and 
the same for  the number of 1's. So if our experimentally determined
values do not fall within the above limits the coin is not fair.
We can look at this from another point of view; namely, what is the
probability that a fair coin will be mistaken for an unfair 
one with the distribution of $(1/3,2/3)$ given n trials on the 
fair coin? For large $n$ the answer is \cite{Vedral2,Cover}

\begin{equation}
p(\mbox{fair}\rightarrow \mbox{unfair}) = e^{-nS_{cl}(uf||f)} \,\,\, ,
\end{equation}
where $S_{cl}(uf||f) = 1/3 \ln 1/3 + 2/3 \ln 2/3 - 1/3 
\ln 1/2 -2/3 \ln 1/2$ is
the Classical Relative Entropy for the 
two distributions. So,
\begin{equation}
p(\mbox{fair}\rightarrow \mbox{unfair}) = 3^n 2^{-\frac{5}{3}n} \,\,\, ,
\end{equation}
which tends exponentially to zero with $n \rightarrow 
\infty$. In fact we see
that already after $\sim 20$ trials the probability of 
mistaking the two 
distributions is vanishingly small, $\le 10^{-10}$.

This result is true, in general, for any two distributions. 
Asymptotically the probability of not distinguishing the 
distributions $P(x)$ and $Q(x)$ after $n$
trials is $e^{-nS_{cl} (P(x)||Q(x))}$,  where 

\begin{equation}
S_{cl}(P(x)||Q(x)) =\sum_i p_i \ln p_i - p_i \ln q_i   
\label{Crelent} 
\end{equation}
(this statement is sometimes called
Sanov's theorem \cite{Cover}). 
To generalize this to quantum theory, we need a means of
generating probability distributions from two quantum
states $\sigma$ and $\rho$. This is accomplished by
introducing a general measurement $E_i^{\dagger}\sum_i E_i= {{\sf 1 \hspace{-0.3ex} \rule{0.1ex}{1.52ex}\rule[-.01ex]{0.3ex}{0.1ex}}}$. So, the probabilities are
given by

\begin{eqnarray}
p_i & = & tr (E_i^{\dagger}E_i \rho) \nonumber \\
q_i & = & tr (E_i^{\dagger}E_i \sigma)  \,\,\,\, .
\end{eqnarray}
Now, we can use eq. (\ref{Crelent}) to distinguish between
$\sigma$ and $\rho$. 
The above is not the most general measurement that we can make, however. In 
general we have $N$ copies of $\sigma$ and $\rho$ in the state 
\begin{eqnarray}
\sigma^N & = & \underbrace{\sigma\otimes \sigma ... \otimes \sigma}_{\mbox{total of N terms}}\\
\rho^N & = & \underbrace{\rho\otimes \rho ... \otimes \rho}_{\mbox{total of N terms}} 
\end{eqnarray}
We may now apply a POVM $\sum_{i} A_i ={{\sf 1 \hspace{-0.3ex} \rule{0.1ex}{1.52ex}\rule[-.01ex]{0.3ex}{0.1ex}}}$ acting on $\sigma^N$ and $\rho^N$. Consequently, we define a new type of relative entropy
\begin{eqnarray}
S_{N}(\sigma ||\rho)  :=  \mbox{sup}_{\mbox{A's}} \{ \frac{1}{N} \sum_{i} tr A_i \sigma^N \ln tr A_i \sigma^N  
 -  tr A_i \sigma^N \ln tr A_i \rho^N \}
\label{lim}
\end{eqnarray}
Now it can be shown that \cite{Donald2}
\begin{equation}
S(\sigma ||\rho) \ge S_{N}
\label{ineq}
\end{equation}
where, as before, 
\begin{equation}
S(\sigma ||\rho) := tr (\sigma \ln \sigma - \sigma\ln\rho)
\end{equation}
is the Quantum Relative Entropy
\cite{Vedral,Vedral2,Lindblad1,Lindblad2,Donald2,Wehrl1} (for the summary of the properties of the Quantum Relative Entropy see \cite{Ohya}). Equality
is achieved in eq. (\ref{ineq}) iff $\sigma$ and $\rho$ commute \cite{Fuchs1}.
However, for any $\sigma$ and $\rho$ it is true that \cite{Petz} 

\[
S(\sigma ||\rho) = \lim_{N\rightarrow \infty} S_{N} \;\; .
\]
In fact, this limit can be achieved by projective measurements which are
independent of $\sigma$ \cite{Hayashi}. 
It is known that if eq. (\ref{Crelent})
is maximized over all general measurements $E$, the upper bound is
given by the quantum relative entropy (see e.g. \cite{Donald2}).
In quantum theory we therefore state a law analogous to Sanov's theorem 
(see also \cite{Vedral2}), 

\noindent
{\bf Theorem 8} (or Quantum Sanov's Theorem). 
The probability of {\bf not} distinguishing 
two quantum states (i.e. density matrices) 
$\sigma$ and $\rho$ after $n$ 
measurements is 

\begin{equation}
p(\rho \rightarrow \sigma) = e^{-nS(\sigma ||\rho)}
\label{prob}
\end{equation}
In fact, as explained before, this bound
is reached asymptotically \cite{Petz}, and the 
measurements achieving this are global projectors independent of 
the state $\sigma$ \cite{Hayashi}. We note that 
the Quantum Sanov Theorem was presented by Donald in
\cite{Donald3} as a definition justified by
properties uniquely characterizing the quantity $e^{-nS(\sigma ||\rho)}$.
The underlying intuition in the above measurement approach and
Donald's approach are basically the same. 
Now the interpretation of the Relative Entropy of Entanglement 
becomes immediately transparent \cite{Vedral2}. The 
probability of mistaking an entangled state $\sigma$ 
for a closest, disentangled state, $\rho$, is 
$e^{-n \times min_{\rho\in {\cal D}} 
S(\sigma,\rho)}=e^{-nE(\sigma)} $.
If the amount of entanglement of $\sigma$ is greater, than it takes 
fewer measurements to distinguish it from a disentangled 
state (or, fixing n, there is a smaller 
probability of confusing it with some disentangled state). 
Let us give an
example. Consider a state $(|00\rangle + |11\rangle)/\sqrt{2}$, 
known to be a maximally entangled state. The closest to it is the
disentangled state $(|00\rangle\langle 00| + |11\rangle\langle 11|)/2$ 
\cite{Vedral}. To distinguish these states it is enough to perform 
projections 
onto $(|00\rangle + |11\rangle)/\sqrt{2}$. If the state that we are 
measuring is the above mixture, then the sequence of results ($1$ 
for a successful projection, and $0$ for an unsuccessful projection) 
will contain on average an equal number of $0$'s and $1$'s. For 
this to be mistaken for the above pure state the sequence has to 
contain all $n$ $1$'s. The probability for that is $2^{-n}$, 
which also comes from using eq. ({\ref{prob}). If, on the
other hand, we performed projections onto the pure
state itself, we would then never confuse it with a mixture,
and from eq. ({\ref{prob}) the probability is seen
to be $e^{-\infty} = 0$. We next apply this
simple idea to obtaining an upper bound to 
the efficiency of any purification procedure.   

\section{Thermodynamics of Entanglement: Purification Procedures}  

There are two ways to produce an upper bound to the efficiency
of any purification procedure. Using condition E3 and the fact
that the Relative Entropy of Entanglement is additive, we can
immediately derive this bound. However, this bound can be 
derived in an entirely different way.
In this section we now abandon conditions E1-E3 and use only 
methods of the previous section to put an upper bound to the
efficiency of purification procedures. In particular, we show that
the entanglement of creation is in general larger than the
entanglement of distillation. This is in contrast with the
situation for pure states where both quantities coincide. 
The Quantum Relative Entropy
is seen to play a distinctive role here, and is singled out as a
`good' generator of a measure of entanglement from among other
suggested candidates.

\subsection{Distinguishability and Purification Procedures}

In the previous section we presented a statistical basis to 
the Relative Entropy of Entanglement by 
considering distinguishability of two (or more) quantum states 
encapsulated in the form of the Quantum Sanov Theorem.  
We now use this Quantum Sanov Theorem to put an upper bound 
on the amount of entanglement that can be distilled 
using any purification procedure. This
line of reasoning follows from the fact that any 
purification scheme can be viewed as a measurement to distinguish 
entangled and disentangled quantum states. Suppose that there exist a 
purification procedure with the following property 

\begin{itemize}
 
\item Initially there are $n$ copies of the state $\sigma$.
If $\sigma$ is entangled, then the end product is $0< m\le n$ singlets 
and $n-m$ states in $\rho \in {\cal D}$.  Otherwise, the final state does 
not contain any entanglement, i.e. $m=0$ (in fact, there is nothing 
special about singlets: the final state can be any other known, 
maximally entangled state because these can be converted
into singlets by applying local unitary operations).
\end{itemize}

\noindent
Note that we can allow the complete knowledge of the state $\sigma$.
We also allow that purification procedures differ for different
states $\sigma$. Perhaps there is a ``universal" purification
procedure independent of the initial state.  
However, in reality, this property is hard to fulfill \cite{Horodecki}. 
At present the best that can be done is to purify a certain class 
of entangled states. (see e.g. \cite{Bennett,Deutsch,Vedral3}). 
The above is therefore an idealization that might never be
achieved. Now, by calculating the upper bound 
on the efficiency of a 
procedure described above we present an absolute 
bound for any particular procedure. 
We ask:
``What is the largest number of singlets that 
can be produced 
(distilled) 
from $n$ pairs in state $\sigma$"? Suppose that we produce $m$ 
pairs. We now project them {\em non--locally} onto the singlet state. 
The procedure will yield positive outcomes ($1$) with certainty 
so long as the state we measure indeed is a
singlet. Suppose that after performing singlet projections onto 
all $m$ particles we get a string of $m$ $1$'s. From this we 
conclude that the final state is a singlet (and therefore the 
initial state $\sigma$ was entangled). However, we could 
have made a mistake. But with what probability? The answer is as follows: 
the largest 
probability of making a wrong inference is $2^{-m} = e^{-m\ln2}$ 
(if the state that we 
were measuring had an overlap with a singlet state of
$1/2$). On the other hand, 
if we were measuring 
$\sigma$ from the very beginning (without performing the 
purification first), then the probability (i.e. the lower 
bound) of the wrong inference would be $e^{-nE(\sigma)}$. 
But, purification procedure might waste some information 
(i.e. it is just a particular way of distinguishing 
entangled from disentangled states, not necessarily 
the best one), so that the following has to hold

\begin{equation}
e^{-nE(\sigma)} \le e^{-m\ln 2} \; ,
\end{equation}
which implies that 

\begin{equation}
nE(\sigma) \ge m \; ,
\end{equation}    
i.e. we cannot obtain more entanglement than is originally 
present. This, of course, is also directly guaranteed by our 
condition E3. The above, however, was a deliberate
exercise in deriving the same result from a different 
perspective, abandoning conditions E1--E3. Therefore the 
measure of entanglement given in eq. (\ref{6a}), when 
$D(\sigma ||\rho) = S(\sigma || \rho)$, can be used to 
provide an upper bound on the efficiency of any 
purification procedure. For Bell diagonal states, Rains \cite{Rains} 
found an upper bound on distillable entanglement using completely 
different methods. It turns out that the bound that he obtains in
this case is identical to the one provided by the relative entropy of
entanglement.

Actually, in the above considerations 
we implicitly assumed that the entanglement of 
$n$ pairs, equivalently prepared in the state $\sigma$, is the same
as $n\times E(\sigma)$. We already indicated that this is a
conjecture with a strongly supported basis in the case 
of the Quantum Relative Entropy. 
Based on the upper bound considerations we can introduce 
the following definition.

\noindent
{\bf Definition 3}. A purification procedure given 
by a local complete positive 
trace preserving map $\sigma \rightarrow \sum V_i 
\sigma V^{\dagger}_i$ is 
defined to be {\em ideal} in terms of 
efficiency iff 

\begin{equation}
\sum tr(\sigma_i) \,\, E( \sigma_i/tr(\sigma_i)) = 
E(\sigma) \,\,\,\,  ,
\end{equation}
where, as usual, $\sigma_i = V_i \sigma V^{\dagger}_i$ 
and $p_i = tr (V_i \sigma V^{\dagger}_i)$ (i.e. a 
the ideal purification is the one where E3 is an
equality rather than an inequality). Notice an apparent formal 
analogy between a purification procedure and the 
Carnot cycle in Thermodynamics. The Carnot cycle 
is the most efficient cycle in Thermodynamics (i.e. 
it yields the greatest 
``useful work to heat" ratio), since it is reversible 
(i.e. it conserves the thermodynamical entropy). 
We would now like to claim that the ideal purification procedure 
is the most efficient purification procedure (i.e. 
it yields the greatest 
number of singlets for a given input state), since 
it is reversible (i.e. it conserves entanglement, 
measured by the minimum of the Quantum Relative 
Entropy over all disentangled states).
Unfortunately
this analogy between the Carnot cycle and 
purification procedures is not exact (it is 
only strictly true for the pure states). 
This is seen when we compare the 
entanglement of creation with the Relative Entropy
of Entanglement. In Theorem 6 we have, in fact, shown
that the entanglement of creation is never smaller 
than the Relative Entropy of Entanglement. 
As an example one can consider Bell diagonal states
for which we can exactly calculate both the entanglement of creation
\cite{Vedral} and the Relative Entropy of Entanglement 
\cite{Bennett1}. 
It turns out that the entanglement of creation is always
strictly larger than the Relative Entropy of Entanglement except
for the limiting cases of maximally entangled Bell states
or of disentangled Bell diagonal states (see Fig. 2 for Werner
states). 
This result leads to the following 

\noindent
{\bf Implication.} In general, the amount of entanglement
that was initially invested in creation of $\sigma$ cannot all be 
recovered (``distilled") by local purification procedures.

\noindent
Therefore, the ideal purification procedure, though most efficient, is 
nevertheless irreversible,
and some of the invested entanglement is lost in the purification
process itself. The solution to this irreversibility lies in the
loss of certain information as can easily be seen from the following analysis.
Suppose we start with an ensemble of $N$ singlets 
and we want to locally create any mixed state 
$\sigma$. Now $\sigma$ can always be written as a mixture 
of pure states $\Psi_1, \Psi_2, ... $ with the
corresponding probabilities $p_1, p_2, ... $
We now use Bennett et al's (de)purification procedure \cite{Bennett}
for pure states
(whose efficiency is governed by the von Neumann
entropy). We convert the first $p_1 \times N$ singlets into
the state $\Psi_1$, second $p_2 \times N$ singlets into the state
$\Psi_2$, and so on... In this way, the whole ensemble is in the
state $\sigma$. But, we have an additional information:
we know exactly that the first $p_1 \times N$ pairs are in the
state $\Psi_1$, second $p_2 \times N$ states are in the state $\Psi_2$,
and so on. This is not the same as being given an initial
ensemble of identically prepared pairs in the state
sigma without any additional information. In this, second, 
case we do not have the additional
information of knowing exactly the state of each of the
pairs. This is why the purification without this knowledge
is less efficient, and hence one expects that the Relative Entropy of Entanglement is smaller than the entanglement of formation.

An open question remains as to whether we can use some
other generator, such as the Bures Metric, to 
give an even more stringent bound on the 
amount of distillable entanglement. 

\subsection{More Than two Subsystems}

We see that the above treatment does not refer to the number 
(or indeed dimensionality) of the entangled systems. This is a desired property
as it makes our measure of entanglement universal. However, in order to
perform minimization in eq. (\ref{measure}) we need to be able to 
define what we mean by a disentangled state of say $N$ particles. 
As pointed out in 
\cite{Vedral2} we believe that this can be done inductively.  Namely, for
two quantum systems, $A_1$ and $A_2$, we define a disentangled state as one 
which can be written as a convex sum of disentangled states of $A_1$ and 
$A_2$ as follows \cite{Vedral,Vedral2}:

\begin{equation}
\rho_{12} = \sum_i p_i \, \rho^{A_1}_i\otimes \rho^{A_2}_i \; ,
\end{equation}
where $\sum_i p_i =1$ and the $p$'s are all positive.
Now, for $N$ entangled systems $A_1, A_2, ... A_N$, the disentangled state is:

\begin{equation}
\rho_{12...N} =  \sum_{\mbox{perm}\{i_1i_2 ... i_N\}}  r_{i_1i_2 ... i_N} 
\rho^{A_{i_1}A_{i_2} ... A_{i_n}}\otimes 
\rho^{A_{i_{n+1}}A_{i_{n+2}} ... A_{i_{N}}} \; ,
\label{perm} 
\end{equation}
where $\sum_{\mbox{perm}\{i_1i_2 ... i_N\}} r_{i_1i_2 ... i_N} = 1$, all 
$r$'s are positive and where $\sum_{\mbox{perm}\{i_1i_2 ... i_N\}}$ is a 
sum over all possible permutations of the set of indices $\{1,2,...,N\}$. 
To clarify this let us see how this looks for $4$ systems:
\begin{eqnarray}
\rho_{1234}  =  & \sum_i & p_i \, \rho^{A_1A_2A_3}_i\otimes \rho^{A_4}_i +  
q_i \, \rho^{A_1A_2A_4}_i\otimes \rho^{A_3}_i \nonumber \\
& + & r_i \, \rho^{A_1A_3A_4}_i\otimes \rho^{A_2}_i + s_i \, 
\rho^{A_2A_3A_4}_i\otimes \rho^{A_1}_i \nonumber \\
& + &  t_i \, \rho^{A_1A_2}_i\otimes \rho^{A_3A_4}_i + u_i \, 
\rho^{A_1A_3}_i\otimes \rho^{A_2A_4}_i \nonumber \\
& + & v_i \, \rho^{A_1A_4}_i\otimes \rho^{A_2A_3}_i
\label{mess} 
\end{eqnarray}
where, as usual, all the probabilities $p_i,q_i, ..., v_i$ are positive 
and add up to unity. The above two equations, at least in principle, define 
the disentangled states for any number of entangled systems. Note that 
this form describes a different situation from the one given 
in eq. (\ref{add}) which refers to a number of pairs shared by Alice
and Bob only. The above 
definition of a disentangled state is justified by extending the idea that
local actions cannot increase the entanglement between two quantum systems 
\cite{Bennett1,Vedral,Vedral2}. In the case of $N$ particles we have $N$
parties (Alice, Bob, Charlie, ... , Wayne) all acting locally on their
systems. The general action that also includes communications can be 
written as \cite{Vedral2}
\begin{equation}
\rho \longrightarrow \!\!\!\!\!\!\! \sum_{i_1,i_2, ... ,I_N} 
\!\!\!\!\!\!\! A_{i_1}\otimes B_{i_2} \otimes ...
\otimes W_{i_N} \, \rho \, A^{\dagger}_{i_1}\otimes B^{\dagger}_{i_2} 
\otimes ...\otimes W^{\dagger}_{i_N}
\label{loc}
\end{equation}  
and it can be easily seen that this action does not alter the form of a 
disentangled state in eqs. (\ref{perm},\ref{mess}). In fact, eq. 
(\ref{perm}) is the most general state invariant {\bf in form} under the 
transformation given by eq. (\ref{loc}). This
can be suggested as a definition of a disentangled state for $N\ge 3$, i.e.
it is the most general state invariant in form under local POVM and classical 
communications. Of course, an alternative to defining a disentangled state
would be 

\begin{equation}
\rho_{12...N} =  \sum_{i}  r_{i} \rho^{A_{1}}_i\otimes \rho^{A_{2}}_i \ldots
\otimes \rho^{A_{N}}_i \;\; ,
\label{last} 
\end{equation}
which means that we do not allow any entanglement in any subset of the
$N$ states. This would be a disentangled state based on some local hidden
variable model. Again we repeat that the particular choice of
a form of disentangled states will depend on the physical
background in our model and there is no absolute sense in which
we can resolve this dichotomy. It should be stressed that for
two particles this free choice does not exist as both pictures
coincide.

\section{Conclusions}

We can look at the entanglement from two different perspectives. One 
insists that local actions cannot increase entanglement and do not 
change it if they are unitary. The other one looks at the way we can
distinguish an entangled state from a disentangled one. In particular,
the following question is asked: what is the probability of confusing 
an entangled state with a disentangled one after 
performing a certain number of
measurements? These two, at first sight different approaches, lead to the
same measure of entanglement. This results in the fact that a purification
procedure can be regarded as a protocol of distinguishing an entangled state
from a disentangled set of states. From this premise we derived the 
upper bound on the 
efficiency of any purification procedure. It turns out that distillable
entanglement is in general smaller than the entanglement of creation.
Our entanglement measure is 
independent on the number of systems and their
dimensionality. This suggests applying it to more than two 
entangled systems
in order to understand multi-particle entanglement. We have shown how 
to compute entanglement efficiently for two spin $1/2$ subsystems using
computational methods. However, a 
closed form for the expression of this entanglement measure is desirable.
However, a closed form for the entanglement of formation has been proposed for two spin $1/2$ particles in \cite{Wootters}.
An interesting problem is to specify all the states that have the 
same amount of entanglement.
We know that all the states that are equivalent up 
to a local unitary transformation have the same amount of entanglement 
(by definition--E2).
However, there are states with the same amount of 
entanglement but which are not 
equivalent up to a local unitary transformation 
(for example one state is pure and the other one is mixed). 
A question for further research is whether they are 
linked by a local complete measurement.
Our work in addition suggest a question of finding a general local map
that preserves the entanglement of a given entangled state.  

\vspace*{0.5cm} 
\noindent
{\bf Acknowledgements}

\noindent
We thank A. Ekert, C.A. Fuchs and P.L. Knight for
useful discussions and comments on the subject of this paper.
We are very grateful to M. J. Donald for communicating an 
alternative proof of Theorem 3 to us and for very useful comments
on the manuscript.
This work was supported by the European Community, the UK Engineering 
and Physical Sciences Research Council, by the Alexander 
von Humboldt Foundation and by the Knight Trust. Part of this
work was completed during the 1997 Elsag-Bailey -- I.S.I. Foundation 
research meeting on quantum computation.

\begin{appendix}
\section{Another proof for the pure state entanglement}

In the following we present a third proof for the value of the 
Relative Entropy of Entanglement for pure states.  
As in the second proof we use the representation of the logarithm
of a density operator in terms of a complex integral as in 
Eq. (\ref{proofnew1})-(\ref{proofnew2}). We would like to know the value of the 
Relative Entropy of Entanglement for a pure state 
$\sigma=|\psi\rangle\langle\psi|$ with 
$|\psi\rangle = \alpha |00\rangle + \beta |11\rangle$. We assume that 
$\rho = \alpha^2 |00\rangle\langle 00| + \beta^2 |11\rangle\langle 11| $
is the closest disentangled state to $\sigma$. Therefore we would have that 
\begin{equation}
	E(\sigma) = S(\sigma||\rho) \;\; .
	\label{app1}
\end{equation}
Assume that we change $\rho$ a little bit, i.e. we have 
\begin{equation}
	\rho_{\lambda} = (1-\lambda) \rho + \lambda \rho^*
	\label{app1a}
\end{equation}
with a small $\lambda$ such that $\rho_{\lambda}$ and $\rho^*$ are 
disentangled. For $\rho$ to be the closest disentangled state to 
$\sigma$ we have to have that 
\begin{equation}
	\frac{d}{d\lambda} S(\sigma || (1-\lambda)\rho + \lambda \rho^*)
	|_{\lambda=0} \ge 0 \;\; .
	\label{app2}
\end{equation}
Using the complex representation of Eq. (\ref{proofnew2}) for the derivative 
of the logarithm we quickly find
\begin{eqnarray}
	\frac{d}{d\lambda} S(\sigma || (1-\lambda)\rho + \lambda \rho^*)
	|_{\lambda=0} 
	&=& - \frac{d}{d\lambda} tr \{ \sigma \ln [ (1-\lambda) \rho +\lambda
	\rho^* ] \}|_{\lambda=0} \nonumber\\[.15cm]
	&=& -\frac{d}{d\lambda} \frac{1}{2\pi i} \oint dz tr
	\{ \sigma \frac{1}{z{{\sf 1 \hspace{-0.3ex} \rule{0.1ex}{1.52ex}\rule[-.01ex]{0.3ex}{0.1ex}}} - \rho_{\lambda}} \} \ln z |_{\lambda=0}
	\nonumber\\[.15cm]
	&=& -\frac{1}{2\pi i} \oint dz tr \{ (\rho^*-\rho) (z{{\sf 1 \hspace{-0.3ex} \rule{0.1ex}{1.52ex}\rule[-.01ex]{0.3ex}{0.1ex}}} - \rho)^{-1}
	\sigma (z{{\sf 1 \hspace{-0.3ex} \rule{0.1ex}{1.52ex}\rule[-.01ex]{0.3ex}{0.1ex}}} - \rho)^{-1} \} \ln z
	\nonumber\\[.15cm]
	&=& 
	1 - tr\{ \rho^* (|00\rangle\langle 00| + |11\rangle\langle 11|
	+ x |00\rangle\langle 11| + x |11\rangle\langle 00| )
	\label{app3}
\end{eqnarray}
where $x=\alpha\beta(\ln \alpha^2 - \ln \beta^2)/(\alpha^2-\beta^2)$ 
and we have used the explicit form of $\sigma$ and $\rho$ together 
with Cauchy's theorem \cite{Churchill}. 
Now we have to show that Eq. (\ref{app3}) is 
always positive. One easily checks that 
\begin{equation}
	x=\alpha\beta(\ln \alpha^2 - \ln \beta^2)/(\alpha^2-\beta^2) \le 1
	\label{app4}
\end{equation}
where the maximum is achieved for $\alpha^2=1/2$. The right hand
side of Eq. (\ref{app3}) can become smallest for $x=1$. For Eq. (\ref{app2})
to be positive we therefore need to show that 
\begin{equation}
	tr\{ \rho^* (|00\rangle\langle 00| + |11\rangle\langle 11|
	+ |00\rangle\langle 11| + |11\rangle\langle 00| ) \le 1
	\label{app5}
\end{equation} 
Using $|\phi^+\rangle = (|00\rangle + |11\rangle)/\sqrt{2}$ 
this follows easily as $\rho^*$ is not entangled and therefore
$\langle \phi^+|\rho^*|\phi^+\rangle \le 1/2$ which immediately 
confirms Eq. \ref{app5}. 
Therefore $\rho$ indeed represents the closest disentangled state
to $\sigma$ and our proof is complete. 

This proof can easily be extended to arbitrary dimensional subsystems
where the maximally entangled states have the form $\sum_n \alpha |nn\rangle$.
In that case the proof becomes more similar to the one presented in 
section II.

\end{appendix}

\newpage

\begin{center}
{\large FIGURE CAPTIONS}
\end{center}

\vspace*{2cm}

Fig1. The set of all density matrices, ${\cal T}$ is represented by the 
outer circle. Its subset, a set of disentangled states ${\cal D}$ is 
represented by the inner circle. A state $ \sigma$ belongs to the 
entangled states, and $ \rho^*$ is the disentangled state that minimizes 
the distance $D( \sigma || \rho)$, thus representing the amount 
of quantum correlations in $\sigma$. State 
$ \rho^*_A \otimes  \rho^*_B$ is obtained by tracing $ \rho^*$ 
over $A$ and $B$. $D( \rho^* || \rho^*_A \otimes  \rho^*_B)$ 
represent the classical part of the correlations in the state $ \sigma$.

\vspace*{2cm}

Fig2. Comparison of the entanglement of creation and the Relative
Entropy of Entanglement for the Werner states (these are are Bell
diagonal states of the form $W=\mbox{diag}(F,(1-F)/3,(1-F)/3,(1-F)/3.$)
One clearly sees that the entanglement of creation is strictly larger
than the Relative Entropy of Entanglement for $0 < F < 1$ 

\newpage

\begin{figure}[b]
\epsfxsize10.0cm
\centerline{\epsfbox{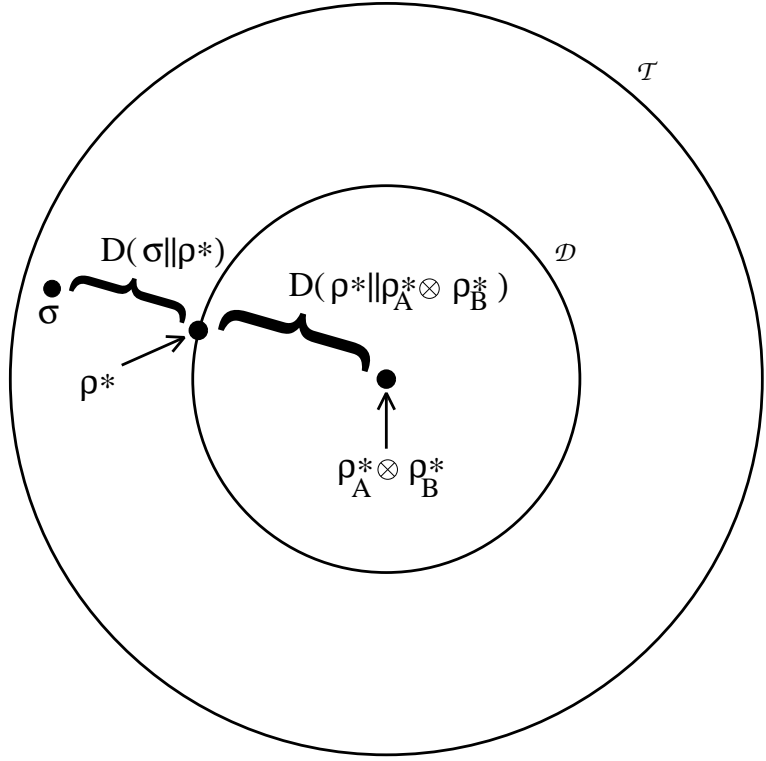}}
\vspace*{1.0cm}
\caption{The set of all density matrices, ${\cal T}$ is represented by the 
outer circle. Its subset, a set of disentangled states ${\cal D}$ is 
represented by the inner circle. A state $ \sigma$ belongs to the 
entangled states, and $ \rho^*$ is the disentangled state that minimizes 
the distance $D( \sigma || \rho)$, thus representing the amount 
of quantum correlations in $\sigma$. State 
$ \rho^*_A \otimes  \rho^*_B$ is obtained by tracing $ \rho^*$ 
over $A$ and $B$. $D( \rho^* || \rho^*_A \otimes  \rho^*_B)$ 
represent the classical part of the correlations in the state $ \sigma$.}
\end{figure}

\newpage

\begin{figure}[b]
\epsfxsize15.0cm
\centerline{\epsfbox{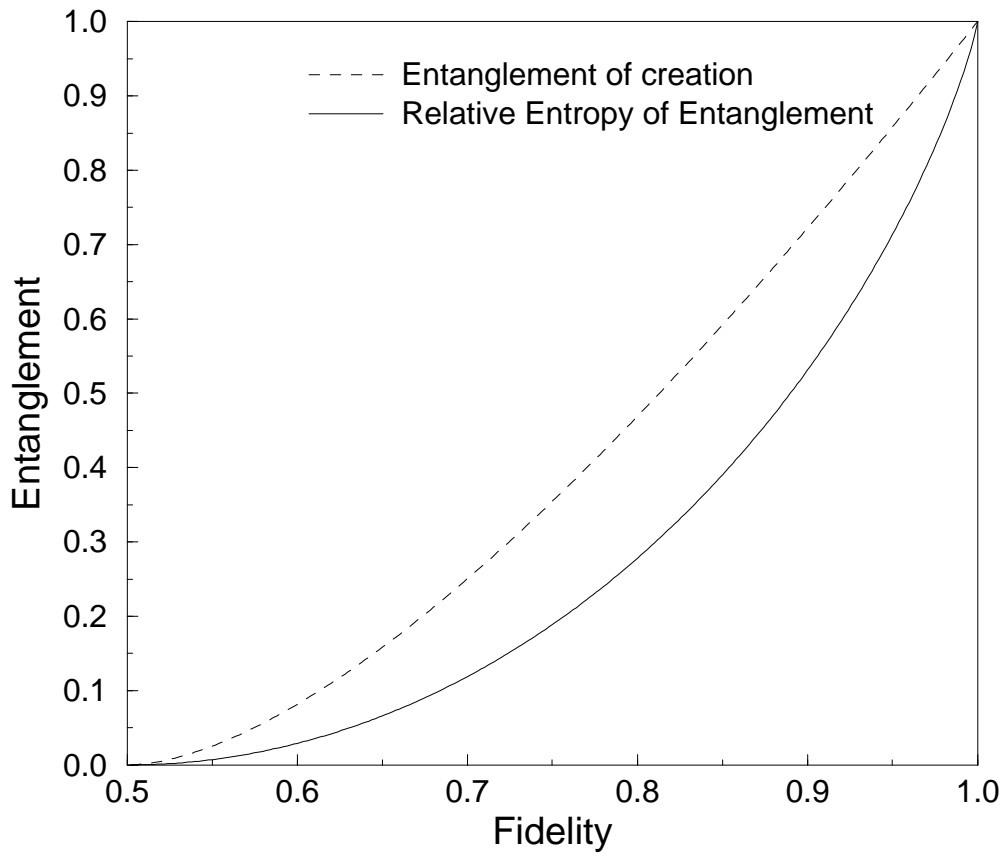}}
\vspace*{1.0cm}
\caption{Comparison of the entanglement of creation and the Relative
Entropy of Entanglement for the Werner states (these are are Bell
diagonal states of the form $W=\mbox{diag}(F,(1-F)/3,(1-F)/3,(1-F)/3.$)
One clearly sees that the entanglement of creation is strictly larger
than the Relative Entropy of Entanglement for $0 < F < 1$}
\end{figure}

\end{document}